\documentclass{pasj01}
\Received{$\langle$reception date$\rangle$}
\Accepted{$\langle$acception date$\rangle$}
\Published{$\langle$publication date$\rangle$}

\begin{document}

\title{Starspots in contact and semi-detached binary systems}
\author{Shinjirou Kouzuma}%
\altaffiltext{}{School of International Liberal Studies, Chukyo University, 101-2 Yagoto-honmachi, Showa-ku, Nagoya, Aichi 466-8666, Japan}
\altaffiltext{}{Departamento de F\'{i}sica de la Tierra y Astrof\'{i}sica, Facultad de Ciencias F\'{i}sicas, Universidad Complutense de Madrid 28040 Madrid, Spain}
\email{skouzuma@lets.chukyo-u.ac.jp,skozuma@ucm.es}


\maketitle

\begin{abstract}
We investigated the statistical properties of both cool and hot starspots in eclipsing binary stars. 
The starspot and binary parameters for contact and semi-detached systems were collected from literature, 
which were determined on the basis of synthetic light-curve analysis. 
We examined associations between these parameters. 
As a result, the cool spots in W-type binaries show properties similar to those of sunspots and starspots generated by dynamos, 
which differs from those of the cool spots in A-type binaries. 
The properties of hot spots also differ between the W- and A-type samples. 
From the physical properties of A- and W-type binaries, 
we infer that mass transfer is a dominant process for forming the hot spots in A-type binaries; 
and both mass transfer and magnetic activity can contribute to the formation of the hot spots in W-type binaries. 
Our results also indicate that the hot-spot size in the A-type sample is correlated with the temperature of spotted stars, orbital period, mass ratio, and fill-out factor. 
\end{abstract}

\section{Introduction}
Starspots are areas on the surfaces of stars, where the temperatures significantly differ from those of the surrounding photospheres. 
Spot activity is closely associated with various stellar phenomena and the inner structure of stars. 
Accordingly, revealing starspot properties is crucial for understanding the stellar activity. 

Two kinds of spots can be defined in terms of the spot temperature: cool and hot spots. 
Cool spots are generally attributed to photospheric magnetic fields. 
The magnetic activity of late-type stars has been explained by the dynamo theory. 
In the dynamo theory, magnetic energy is generated by the conversion of kinetic energy that comes from convection in the outer layer. 
The magnetic activity leads to various stellar activity. 
Flares, for instance, have been interpreted as a result of the release of magnetic energy (e.g. \cite{Shibata2011-LRSP}). 
\citet{Sammis2000-ApJ} reported that the sunspot coverage increases as the energy of the largest flare increases. 
Recent studies discovered correlations between the starspot activity and superflares on solar-type stars \citep{Notsu2013-ApJ,Maehara2017-PASJ}. 
Thus, the spot activity is closely associated with stellar phenomena, as well as the magnetic activity. 

Hot spots have also been believed to be associated with the magnetic activity of stars. 
Solar faculae are often observed on areas which surround cool spots. 
However, as for close binary systems, mass exchange between component stars is another plausible scenario for creating hot spots, 
since transferred material collides on the surface of a component star. 
These hot spots are expected to appear in contact and semi-detached systems 
because such systems contain at least one star that filling its Roche lobe. 

Binary stars can have starspots whose properties differ from those of sunspots, 
although similarities between starspots and sunspots have been reported. 
Sunspots cover at most a few percent of the Sun's surface area, and typically live for hours to months \citep{Solanki2003-AARv}. 
By contrast, gigantic starspots covering up to $20\%$ were found in RS CVn-type binary systems (e.g. \cite{Strassmeier1999-AA225}), and 
a long-lived ($\sim 11$ yr) polar spot was also detected \citep{Vogt1999-ApJS}. 
Furthermore, \citet{Hussain2002-AN} concluded that spots on tidally locked binary systems live longer than spots on single main-sequence stars. 
Accordingly, the starspots in binary systems should have the properties that differ from those of the spots on single stars as well as sunspots. 

Starspot activity leads to a distorted light-curve in many cases. 
Eclipsing binaries, as well as single stars, can exhibit light curves distorted by the presence of starspots. 
The O'Connell effect \citep{OConnell1951-PRCO,Milone1968-AJ} is characterized by an asymmetric light-curve with unequal out-of-eclipse maxima. 
This phenomenon has been often explained by the presence of starspots. 
Many authors have modeled distorted light-curves assuming that cool or hot spots are present on the stellar surfaces. 
Other several techniques are also available to detect starspots: 
Doppler imaging \citep{Vogt1983-PASP}, molecular bands modeling \citep{Vogt1979-PASP,Huenemoerder1989-AJ}, 
eclipse mapping \citep{Collier1997-MNRAS,Lister2001-MNRAS}, and gravitational microlensing \citep{Heyrovsky2000-ApJ}.
The synthetic light-curve analysis, nevertheless, is a relatively easy method to estimate spot parameters. 
In practice, the spot parameters for a number of binary stars have been determined by previous studies. 

RS CVn-type stars, which are close detached binaries with magnetically active components, have been a great target for studying starspots in binary systems. 
The reason is that they are expected to have starspots generated by their strong magnetic activity. 
However, unlike the starspot properties in RS CVn-type stars, those in contact and semi-detached binaries are poorly understood. 
Although starspot parameters have been determined for various close binary systems on the basis of the synthetic light-curve analysis, 
their statistical properties are less known. 

This paper presents the statistical properties of both cool and hot spots in contact and semi-detached systems. 
Section \ref{Sample} introduces the starspot and binary parameters we collected. 
We examine associations between the astrophysical parameters in section \ref{Properties} and discuss their properties in section \ref{Discussion}. 
Section \ref{Conclusion} summarizes our results.

\begin{table}[tbp]
\caption{Statistics of collected binary systems with spots. 
\label{Num-spotted}}
\begin{center}
 \begin{tabular}{cccccc}
  \hline
          & W-type & A-type & SD1 & SD2 & Total\\
  \hline
Cool spot\footnotemark[$*$]                   & 52 (27) & 32 (21) & 3 (2) & 15 (4) & 102 (54)   \\
\multicolumn{1}{r}{L/M\footnotemark[$\dag$]}   & 11/41   & 13/19   & 1/2   & 11/4   & 36/66 \\
\multicolumn{1}{r}{C/H\footnotemark[$\ddag$]}  & 39/13   & 21/11   & 1/2   & 11/4   & 72/30 \\
Hot spot\footnotemark[$*$]                     & 15 (7)  & 16 (12) & 6 (0) & 8 (2)  & 45 (21)  \\
\multicolumn{1}{r}{L/M\footnotemark[$\dag$]}   & 5/10    & 8/8     & 6/0   & 1/7    & 20/25 \\
\multicolumn{1}{r}{C/H\footnotemark[$\ddag$]}  & 10/5    & 8/8     & 6/0   & 1/7    & 25/20 \\
  \hline
 \end{tabular}
\begin{tabnote}     
{\hbox to 0pt{
\parbox{78mm}{
\footnotesize 
\par
\hangindent6pt\noindent
\hbox to 6pt{\footnotemark[$*$\hss]}\unskip
The numbers in parentheses represent the numbers of systems whose parameters were determined on the basis of spectroscopic mass-ratio. 

\par
\hangindent6pt\noindent
\hbox to 6pt{\footnotemark[$\dag$\hss]}\unskip
The symbols L and M denote that the less-massive and more-massive component stars have a starspot respectively. 
\par
\hangindent6pt\noindent
\hbox to 6pt{\footnotemark[$\ddag$\hss]}\unskip
The symbols C and H denote that the cooler and hotter component stars have a starspot respectively. 
}\hss}}
\end{tabnote}
\end{center}
\end{table}
\section{Sample of spotted binary systems}\label{Sample}
We collected the starspot parameters of eclipsing binaries from literature, together with their physical parameters. 
These parameters were determined by synthetic light-curve analysis based on methods such as \citet{Wilson1971-ApJ} and \citet{Djurasevic1992-ApSS}. 
We selected systems whose parameters were determined on the basis of multi-color light curves.
These collected binaries exhibited distorted light-curves. 
The authors, who studied the collected binaries, analyzed the light curves assuming the distortions were due to the presence of starspots. 
When the solution from a light-curve analysis indicates the presence of more than one spot, we selected the parameters of the largest spot. 
When no spectroscopic data are available, 
the determination of mass ratio tends to deteriorate for binary systems with low inclination. 
\citet{Maceroni1996-AA} compared the photometrically- and spectroscopically-determined mass-ratios, 
in which they used a sample of binary systems with inclinations larger than about 60 deg. 
Their comparison showed that the photometric mass-ratios are similar to the spectroscopic ones. 
Taking their result into account, we also excluded systems having both inclinations smaller than $75$ deg and photometrically-determined mass-ratios. 
Finally, we extracted cool and hot spot samples for 102 and 45 binary systems respectively. 
Note that spots only for contact and semi-detached systems were extracted. 

Table \ref{Num-spotted} summarizes statistics for the cool and hot spots. 
Both W- and A-types are the subtypes of contact binary, which are based on \citet{Binnendijk1970-VA}. 
The symbols SD1 and SD2 denote semi-detached binary with the more-massive and less-massive component filling its Roche lobe respectively. 

Tables \ref{CS-binaries} and \ref{HS-binaries} list the astrophysical parameters for the collected binaries with cool and hot spots respectively. 
Throughout this paper, colatitude ($\theta$) runs from $0^\circ$ at the $+z$ pole to $180^\circ$ at the $-z$ pole, 
and longitude ($\phi$) runs counterclockwise from $0^\circ$ to $360^\circ$ as seen from gaboveh ($+z$).
If the errors for a parameter are estimated in the relevant references, they are added as error bars to the figures of this paper. 
Almost all starspots in tables \ref{CS-binaries} and \ref{HS-binaries} have angular radii larger than $10^\circ$. 
Starspots smaller than $10^\circ$ may be difficult to be detected with synthetic light-curve analysis 
because small starspots hardly distort light curves. 
Note that a starspot detected with light-curve modeling can be a group of smaller spots.

\begin{figure*}[tbp]
\begin{center}
\includegraphics[width=55mm]{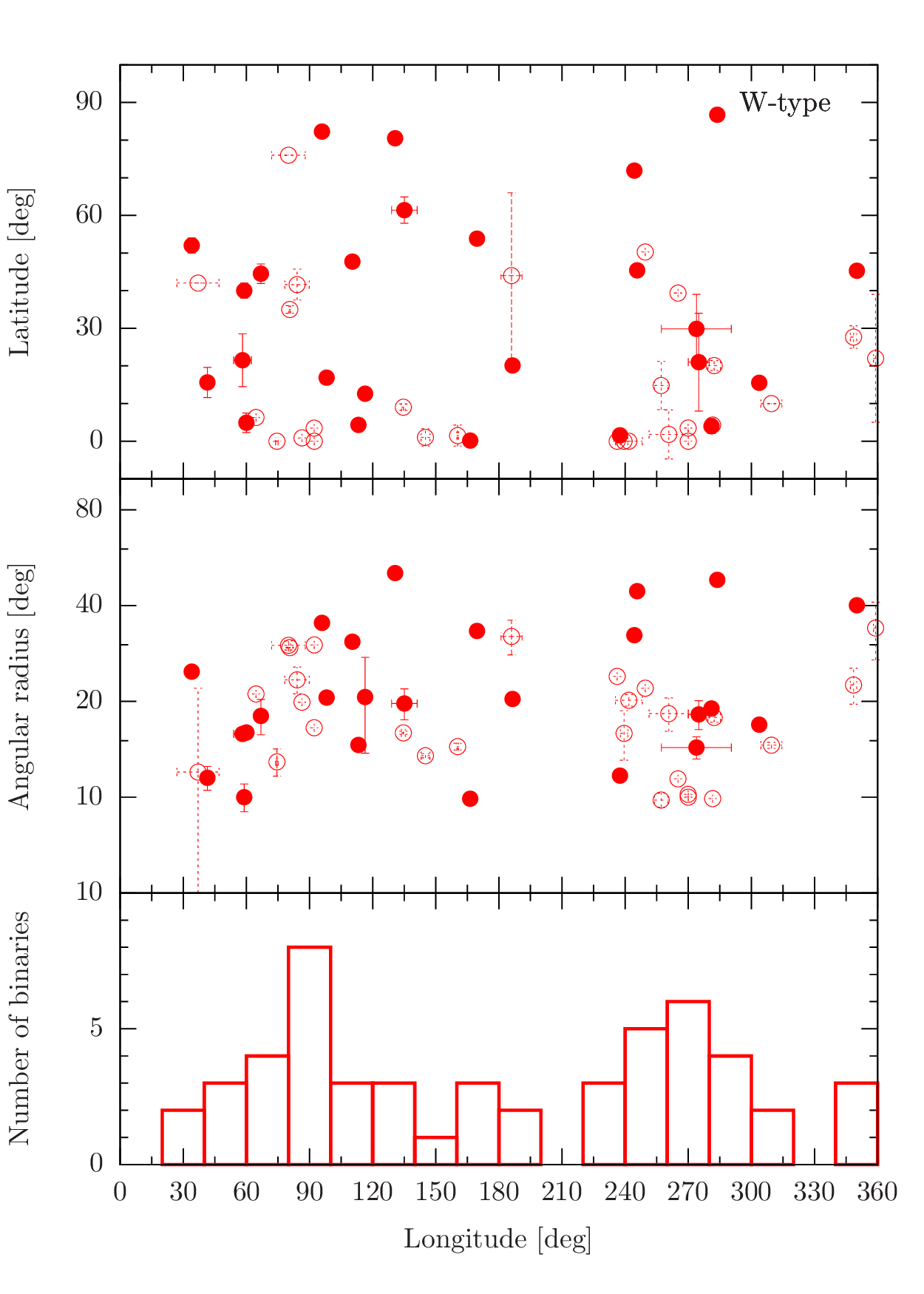}
\includegraphics[width=55mm]{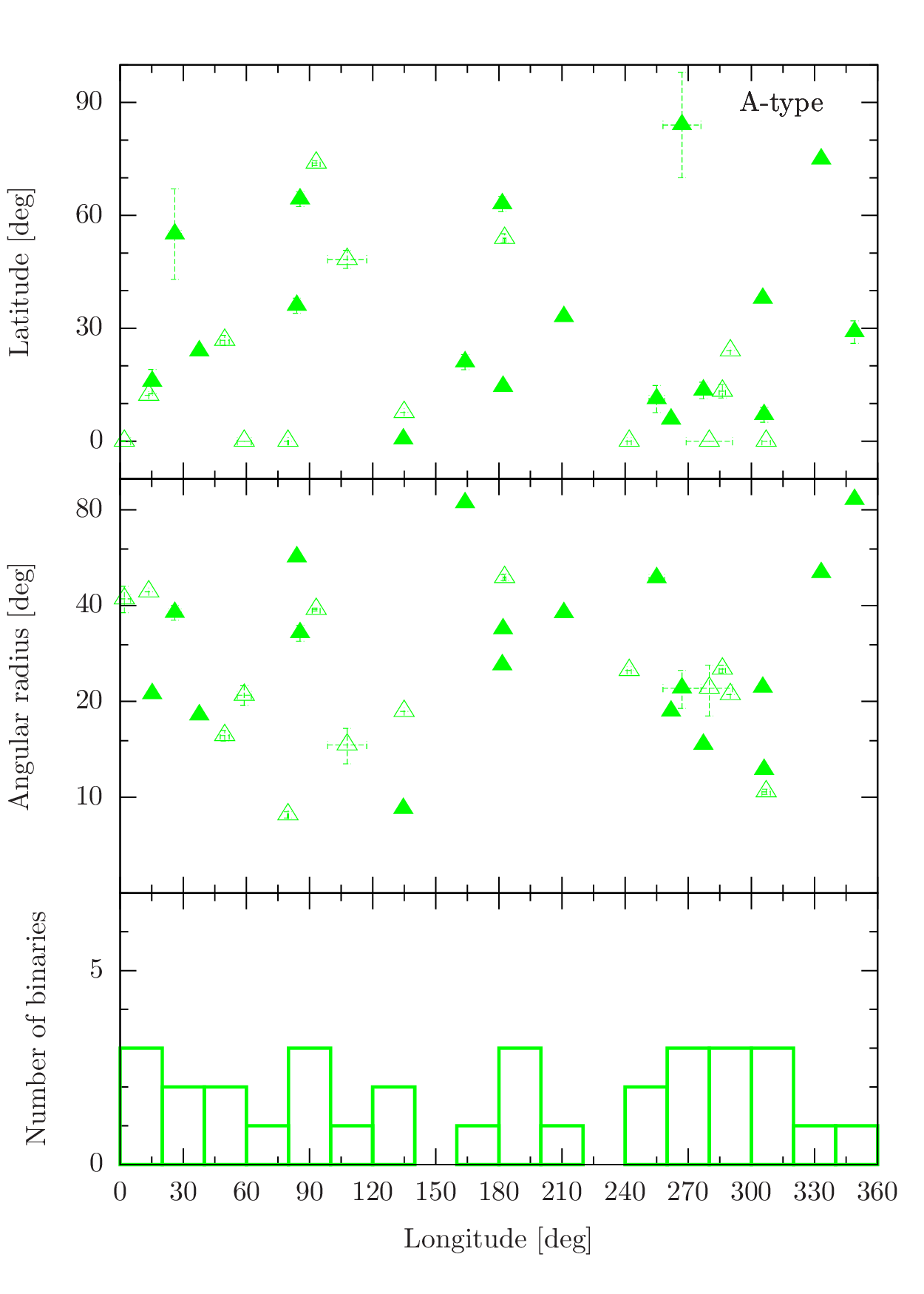}
\includegraphics[width=55mm]{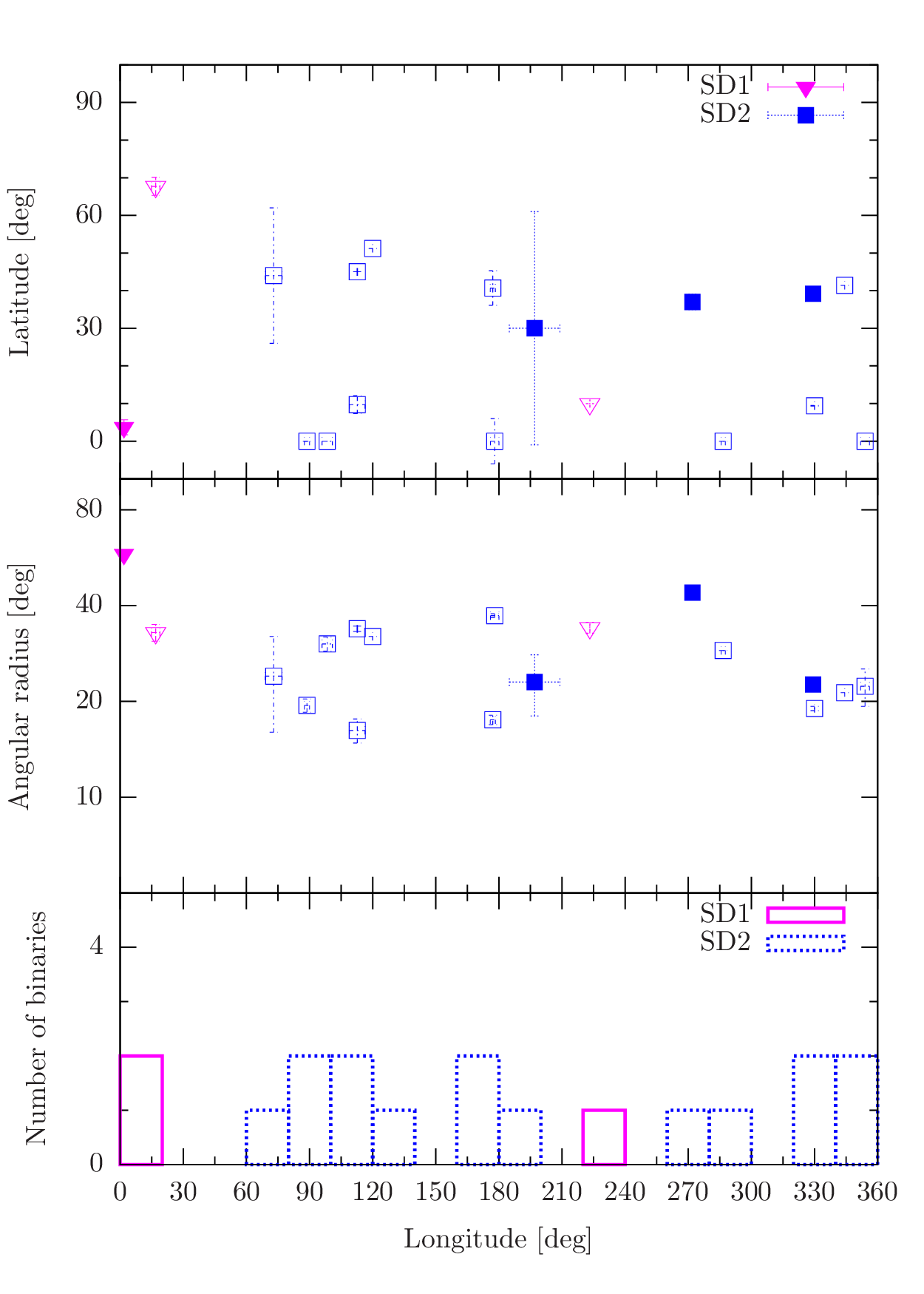}
\end{center}
\caption{
Longitudinal distributions of cool spots in contact and semi-detached binaries: 
cool spot distribution on the stellar surfaces (top panel), 
angular radius as a function of spot longitude (middle panel), and 
histograms of spot longitude (bottom panel). 
The solid and open symbols represent binary systems whose mass ratios were determined on the basis of spectroscopic and photometric data respectively. 
The binaries with fixed spot parameters are represented by open symbols, even if they have spectroscopic mass-ratio. 
\label{Lon-Rad-Lat-CS}}
\end{figure*}

\begin{figure}[tbp]
\begin{center}
\includegraphics[width=80mm]{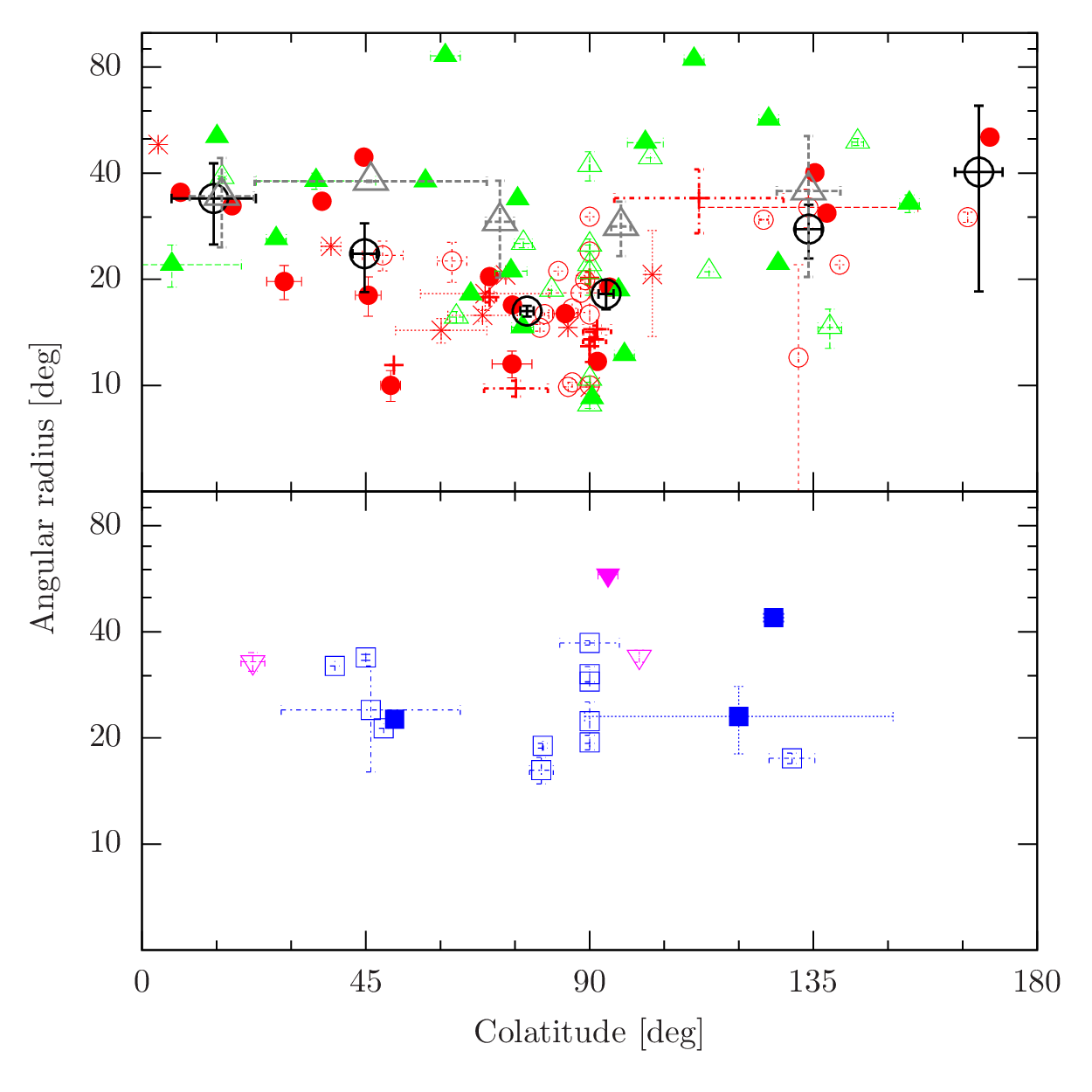}
\end{center}
\caption{Colatitude versus angular radius for cool spots. 
The larger open circles and open triangles represent mean values that are appropriately divided into each bin. 
The asterisk and plus symbols represent the W-type binaries with and without spectroscopic mass-ratio respectively, 
and these are located below the 5-$\sigma$ line in figure \ref{Tstar-DeltaT-CS. 
}
Other symbols are the same as in figure \ref{Lon-Rad-Lat-CS}. 
\label{Lat-Rad-CS}}
\end{figure}

\section{Statistical properties of starspots}\label{Properties}
\subsection{Cool spot}\label{CS-PRO}
\subsubsection{Position}\label{CS-Position}
Figure \ref{Lon-Rad-Lat-CS} shows the scatter plots of longitude versus latitude and longitude versus the angular radius of spots, 
together with histograms of the number of binaries plotted as a function of longitude. 
In this paper, latitude ($\lambda$) is measured from $0^\circ$ at the stellar equator (orbital plane) to $90^\circ$ towards both the north and south poles. 
The histogram for the W-type sample shows a bimodal shape, with two peaks around $\phi=90^\circ$ and $\phi=270^\circ$. 
\citet{Jetsu1991-LNP} first discovered flip-flop phenomenon in FK Comae, which has also been detected in RS CVn-type binaries (e.g. \cite{Berdyugina1998-AA}; \cite{Berdyugina1999-AA}). 
This phenomenon is a switch of the activity between two active longitudes which are separated by $180^\circ$ on average. 
The longitudinal distance between the two peaks also is around $180^\circ$, 
which agrees with the separation appears in the flip-flop phenomenon. 
Meanwhile, \citet{Jeffers2005-MNRAS} demonstrated that the preferred longitudes of spot emergence can be reproduced by a randomly-distributed spot model. 
Their model could explain the longitudinal distribution of our W-type sample. 
However, the W-type sample also shows that the high-latitude ($\lambda>$70$^\circ$) spots in five W-type systems were present at around $\phi=90^\circ$ and $\phi=270^\circ$ (see the top-left panel in figure \ref{Lon-Rad-Lat-CS}) 
and their angular radii are relatively large ($\alpha>30^\circ$). 
The Doppler imaging technique has found many large spots at high latitudes on magnetically active stars (e.g. \cite{Vogt1983-PASP,Hendry2000-ApJ}). 
The five W-type systems also have large and high-latitude spots, which agrees with the result from the Doppler imaging. 
This also supports that the spot activity of W-type systems is strong around $\phi=90^\circ$ and $\phi=270^\circ$. 

The flip-flop phenomenon has been explained by dynamo models (\cite{Moss2004-MNRAS}; \cite{Korhonen2005-AA}). 
The component stars of W-type binaries generally have G--K spectra \citep{Webbink2003-ASPC}. 
Because such stars have convective envelopes, their internal dynamos can generate magnetic fields. 
Hence, when the magnetic activity forms cool spots, the flip-flop phenomenon is expected to arise on the spotted stars in W-type systems. 
If the flip-flop phenomenon indeed arises in W-type binaries, 
the properties of the W-type sample indicate that active longitudes of W-type binaries tend to be present around the vertical sides of the spotted star facing to another component. 

Contrary to the W-type sample, other samples have few characteristics. 
No clear characteristics are found in the semi-detached samples, which may be due to low statistics. 
However, it seems that A-type sample binaries tend to have hot spots at $\phi \sim270^\circ$.

Another notable feature includes the latitudinal dependence of the spot size in figure \ref{Lat-Rad-CS}. 
The spots of W-type sample binaries tend to be larger for spot positions close to the poles of spotted stars. 
To examine the statistical significance of the association, 
we computed both the Pearson correlation coefficient ($r_\textnormal{\scriptsize p}$) and the Spearman rank-correlation coefficient ($r_\textnormal{\scriptsize s}$). 
The $r_\textnormal{\scriptsize p}$ and $r_\textnormal{\scriptsize s}$ values are suitable for evaluating the linear and monotonic relations between two variables, respectively. 
The calculated coefficients are 
$r_\textnormal{\scriptsize p}=-$0.638 with $p<0.001$ and 
$r_\textnormal{\scriptsize s}=-$0.396 with $p=0.015$ in the range $0^\circ < \theta< 90^\circ$, and 
$r_\textnormal{\scriptsize p}=$0.707 with $p<0.001$ and 
$r_\textnormal{\scriptsize s}=$0.569 with $p=0.007$ in the range $90^\circ < \theta< 180^\circ$. 
The values of $p$ are $p$-values which are calculated on the basis of null hypothesis that there is no linear or monotonic relations between two variables. 
These values indicate that the spot size increases with increasing latitude. 
This tendency is similar to the sunspot property that the cycle-integrated sunspot area is positively correlated with the mean latitude 
\citep{Li2003-SoPh,Solanki2008-AA,Ivanov2016-GeAe}. 
Although the A-type binaries also show a similar association at least below $\lambda \sim 60^\circ$ 
($r_\textnormal{\scriptsize p}=-$0.415 with $p=0.110$ and 
$r_\textnormal{\scriptsize s}=-$0.356 with $p=0.175$ in the range $30^\circ < \theta< 90^\circ$, and 
$r_\textnormal{\scriptsize p}=$0.334 with $p=0.191$ and 
$r_\textnormal{\scriptsize s}=$0.430 with $p=0.085$ in the range $90^\circ < \theta< 150^\circ$), 
these correlations are weaker than those of the W-type sample. 
The semi-detached samples seem to show no significant association between the latitude and the angular radius.

\begin{figure}[tbp]
\begin{center}
\includegraphics[width=80mm]{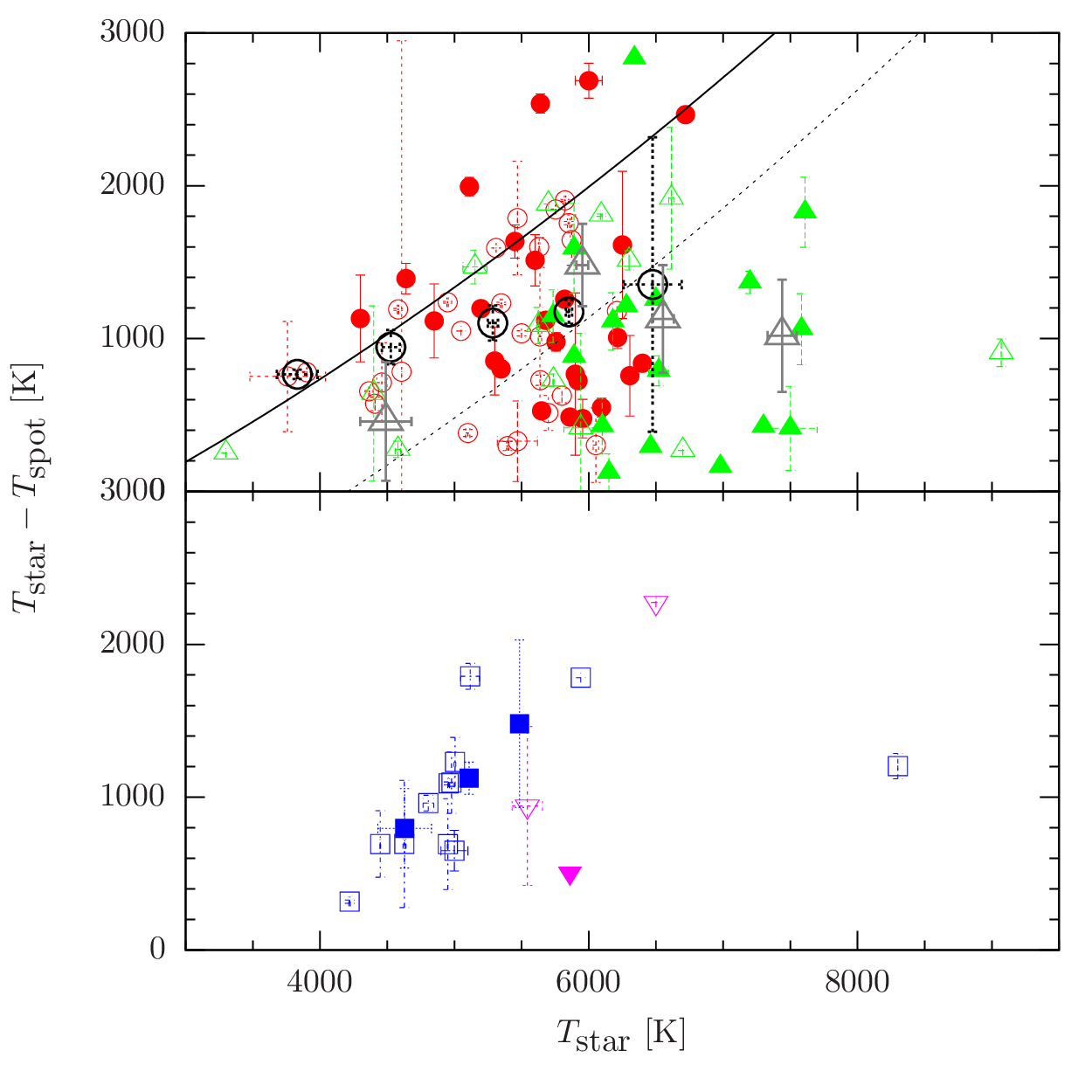}
\end{center}
\caption{
Photosphere temperature of spotted star versus temperature difference between photosphere and cool spot. 
Symbols are the same as in figures \ref{Lon-Rad-Lat-CS} and \ref{Lat-Rad-CS}. 
}
\label{Tstar-DeltaT-CS}
\end{figure}
\subsubsection{Temperature}\label{CS-Temperature}
\citet{Berdyugina2005-LRSP} presented a representative sample of starspot temperatures for active dwarfs, giants and subgiants. 
The temperature difference between spots and the photosphere for the sample decreases from about 2000 K in G0 stars to 200 K in M4 stars. 
In figure \ref{Tstar-DeltaT-CS}, we depict the scatter plot between the photosphere temperature and the temperature difference, 
together with a second-order polynomial fit to the data for the sample of \citet{Berdyugina2005-LRSP} except for EK Dra. 
The polynomial fit line is 
\begin{eqnarray}
\Delta T &\equiv& (T_\textnormal{\scriptsize phot}-T_\textnormal{\scriptsize spot}) \nonumber \\
&=&2.89\times10^{-5}T_\textnormal{\scriptsize phot}^2+0.34T_\textnormal{\scriptsize phot}-1088, 
\end{eqnarray}
where $T_\textnormal{\scriptsize phot}$ and $T_\textnormal{\scriptsize spot}$ are photosphere and spot temperatures, respectively. 
The root mean square of the residuals of the fit is 171 K. 
Although our W-type sample also shows a positive correlation 
($r_\textnormal{\scriptsize p}=$0.209 with $p=$0.137 and $r_\textnormal{\scriptsize s}=$0.072 with $p=$0.612) 
similar to that of \citet{Berdyugina2005-LRSP}, 
its distribution is significantly spread out downward. 
The broken line in figure \ref{Tstar-DeltaT-CS} is a line 5-$\sigma$ lower than the fitting one. 
Several W-type binaries are located in the area below the 5-$\sigma$ line, that is, they have relatively small temperature differences. 
In the sample of \citet{Berdyugina2005-LRSP}, EK Dra had a large temperature ($T \sim 5900$ K) and a small temperature-difference ($\Delta T=500$--$1050$ K), 
and they surmised that active late F-type stars possess spots with dominating penumbra. 
Thus, starspots located below the 5-$\sigma$ line also may be dominated by penumbra. 

W- and A-type sample binaries are roughly separated by $T \sim 6000$ K. 
This is reasonable because W- and A-type systems generally have G--K and A--F spectra, respectively \citep{Webbink2003-ASPC}. 
Three A-type systems have temperatures lower than 5000 K: 
UCAC4 436-062932 (UCAC4 436), 1SWASP J074658.62+224448.5 (SW J074), and 1SWASP J075102.16+342405.3 (SW J075). 
However, it is arguable that the subtypes of these systems are accurate. 
Another author who analyzed UCAC4 436 classified the binary as a W-type system \citep{Zhou2016-NewA}. 
SW J074 and SW J075 have short orbital-periods compared with other A-type systems (see also section \ref{CS-P}). 
Accordingly, the subtypes of these binaries should be further investigated. 
The A-type sample shows no significant association, even if these suspicious binaries are excluded. 

The spotted stars of SD2 sample binaries generally have temperatures lower than $6000$ K, 
which are relatively low compared to those of SD1 sample binaries. 
In addition, the SD2 sample has a positive correlation 
($r_\textnormal{\scriptsize p}=$0.836 with $p<$0.001 and 
$r_\textnormal{\scriptsize s}=$0.815 with $p<$0.001), 
as well as the W-type sample.

\begin{figure}[tpb]
\begin{center}
\includegraphics[width=80mm]{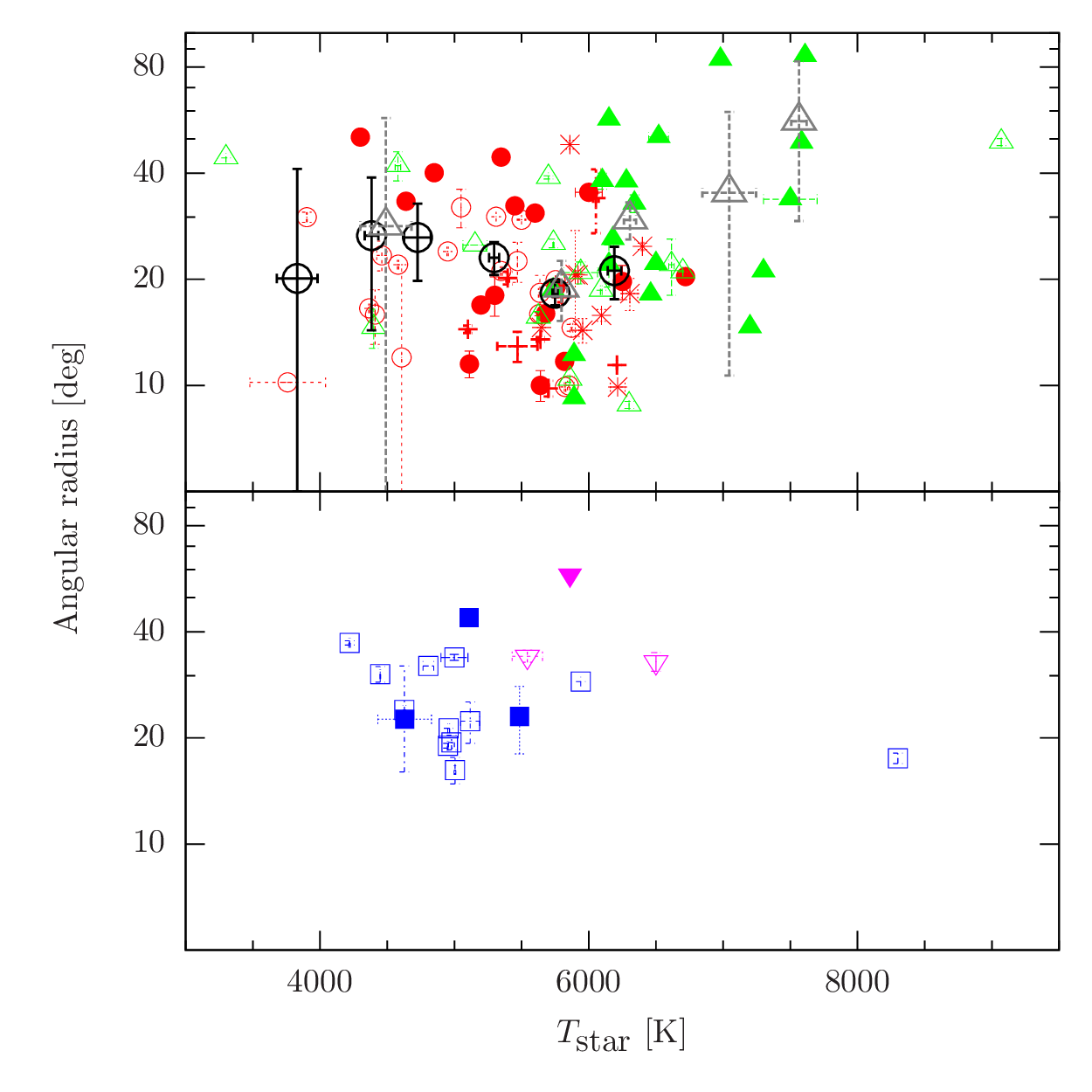}
\end{center}
\caption{
Photosphere temperature of spotted star versus angular radius of cool spot. 
Symbols are the same as in figures \ref{Lon-Rad-Lat-CS} and \ref{Lat-Rad-CS}. 
}
\label{Tstar-Rad-CS}
\end{figure}

Figure \ref{Tstar-Rad-CS} shows the relation between the spotted-star temperature and the angular radius of the starspot. 
\citet{Berdyugina2005-LRSP} presented a relation between photosphere temperature and spot coverage. 
Their relation had a downward parabolic association with a peak around $T=4500$ K. 
Our W-type sample also shows a similar downward parabolic association. 
Meanwhile, the A-type sample has a positive correlation above $T\sim5500$ K 
($r_\textnormal{\scriptsize p}=$0.469 with $p=0.012$ and 
$r_\textnormal{\scriptsize s}=$0.437 with $p=0.020$).

\begin{figure}[tpb]
\begin{center}
\includegraphics[width=80mm]{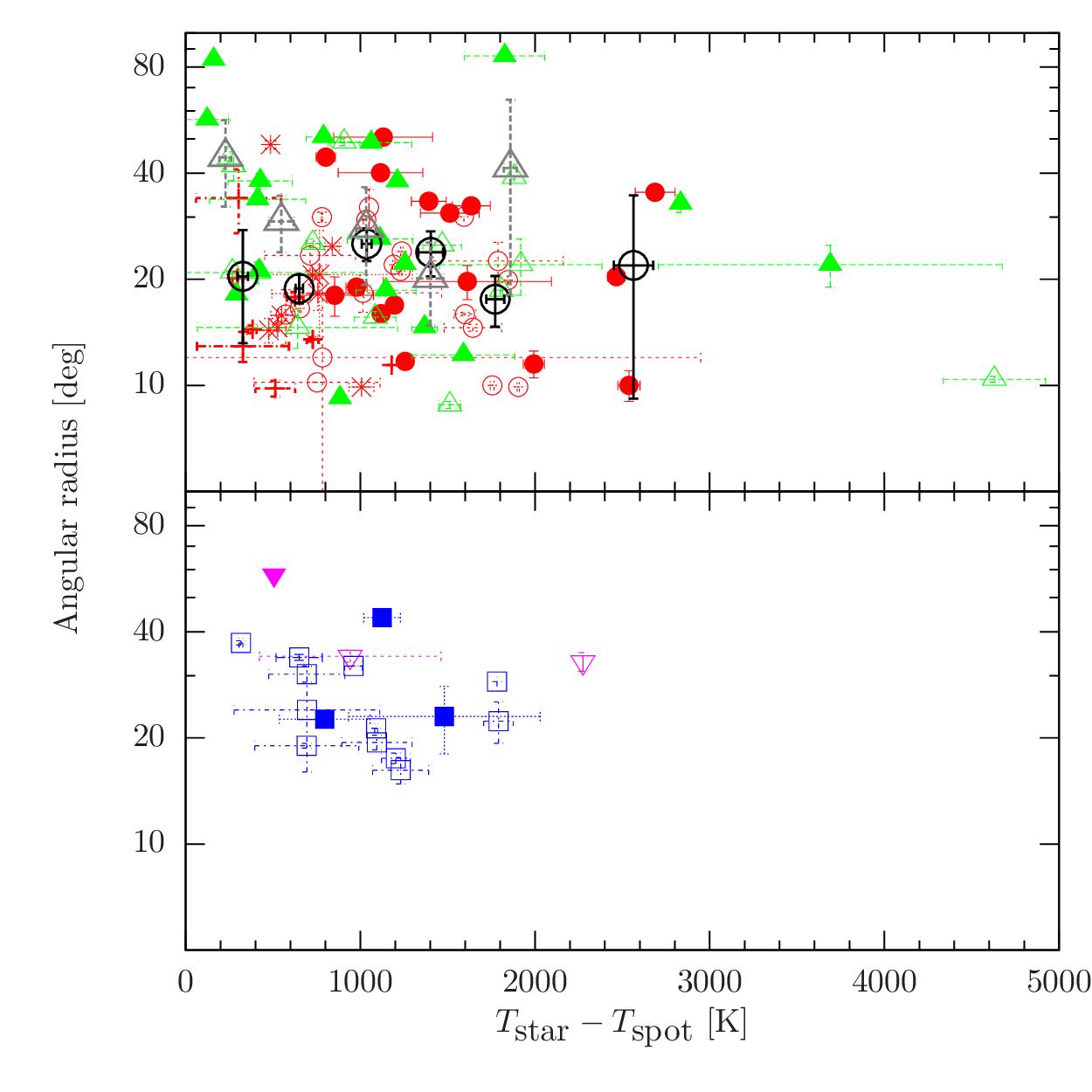}
\end{center}
\caption{
Temperature difference versus angular radius of cool spot. 
Symbols are the same as in figures \ref{Lon-Rad-Lat-CS} and \ref{Lat-Rad-CS}. 
}
\label{DeltaT-Rad-CS}
\end{figure}

In figure \ref{DeltaT-Rad-CS}, we depict the scatter plot between the temperature difference and the spot size. 
Although the W-type sample binaries with spectroscopic mass-ratios appear to have a negative association, 
this is weak 
($r_\textnormal{\scriptsize p}=-$0.256 with $p=0.322$ and $r_\textnormal{s}=-$0.221 with $p=0.395$). 
This tendency agrees with previous studies \citep{Bouvier1989-AA,Strassmeier1992-ASPC}. 
By contrast, the A-type sample shows a negative association 
($r_\textnormal{\scriptsize p}=-$0.253 with $p=0.162$ and $r_\textnormal{s}=-$0.291 with $p=0.106$). 
The cool spot in DU Boo is considerably large ($\alpha=86^\circ$) and large temperature difference ($\Delta T=1826$ K) compared with the cool spots in other A-type binaries. 
\citet{Djurasevic2013-AJ} concluded that the cool spot in DU Boo is due to the exchange of thermal energy. 
Hence, the spot formation of DU Boo may differ from that of other A-type binaries. 
When DU Boo is excluded from the A-type sample, the positive association becomes significant 
($r_\textnormal{\scriptsize p}=-$0.367 with $p=0.042$ and $r_\textnormal{s}=-$0.386 with $p=0.032$).

\begin{figure}[tbp]
\begin{center}
\includegraphics[width=80mm]{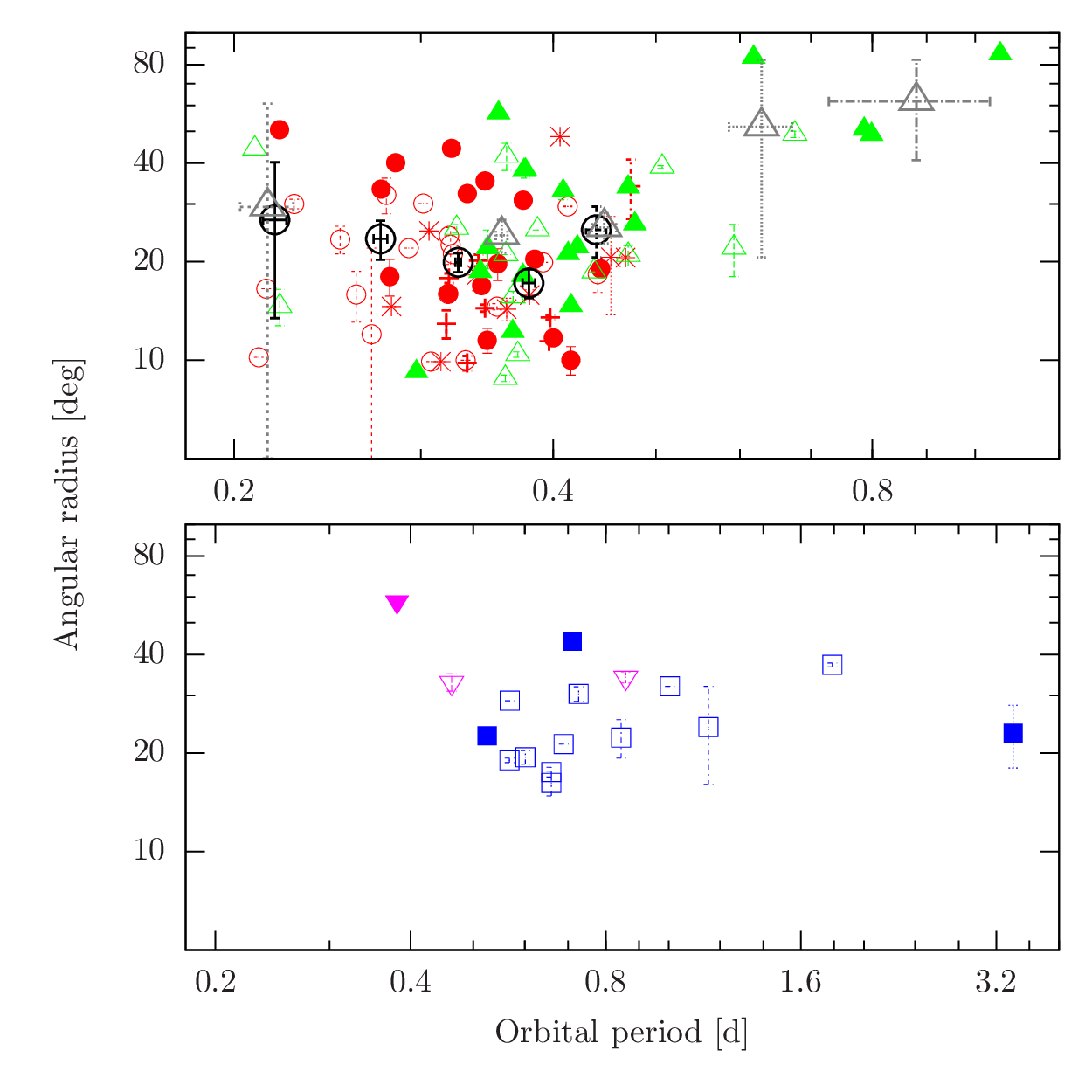}
\end{center}
\caption{
Orbital period versus angular radius of cool spot. 
Symbols are the same as in figures \ref{Lon-Rad-Lat-CS} and \ref{Lat-Rad-CS}. 
}
\label{P-Rad-CS}
\end{figure}
\subsubsection{Orbital period}\label{CS-P}
The relation between the orbital period and the spot size are shown in figure \ref{P-Rad-CS}. 
The contact binary samples, as a whole, has an upward parabolic association with a peak around $P=0.3$--$0.4$ d. 
In other words, the angular radii of W-type sample binaries decrease with increasing orbital period, 
whereas those of the A-type sample binaries increase with increasing orbital period. 

The sample of contact binaries from \citet{Gazeas2006-MNRAS} exhibits that 
all systems with $P<0.3$ d are of W-type and all with $P>0.6$ d are of A-type. 
Our samples include three A-type systems with $P<0.3$ d: 
TZ Boo, SW J074, and SW J075. 
Of these binaries, the latter two are the same as the A-type systems with considerably low-temperature spotted-stars mentioned in section \ref{CS-Temperature}. 
The orbital period of TZ Boo is extremely close to $0.3$ d ($P=0.29716$ d), and the classification into A-type system is possible. 
However, \citet{Christopoulou2011-AJ} pointed out that TZ Boo is one of the most puzzling W UMa systems and that their classification is still unclear. 

The orbital-period distribution of contact binaries is known to have a sharp cut-off around $P\sim0.22$ d \citep{Rucinski2007-MNRAS}. 
Whereas the two W-type systems with $P<0.22$ d (namely 2MASS 02272637+1156494 and 1SWASP J015100.23-100524.2) have small spots of $\alpha<20^\circ$, 
other two systems with $P\sim0.22$ d (namely CC Com and V1104 Her) have relatively large spots ($\alpha>30^\circ$). 
\citet{Rucinski1992-AJ960} explained the period cut-off by the stars reaching full convective limit. 
A recent study demonstrated that the short-period limit is due to the instability of mass-transfer; 
it occurs when the primaries of the initially detached binaries fill their Roche lobes \citep{Jiang2012-MNRAS}. 
Although its exact mechanism is still open to question, 
contact binaries with periods close to the limit are expected to have properties differing from those of other contact binaries. 
Therefore, this tendency indicates that contact binaries with $P<0.22$ d tend to have small spots, unlike those with periods close to but larger than the period limit.

\begin{figure}[tbp]
\begin{center}
\includegraphics[width=80mm]{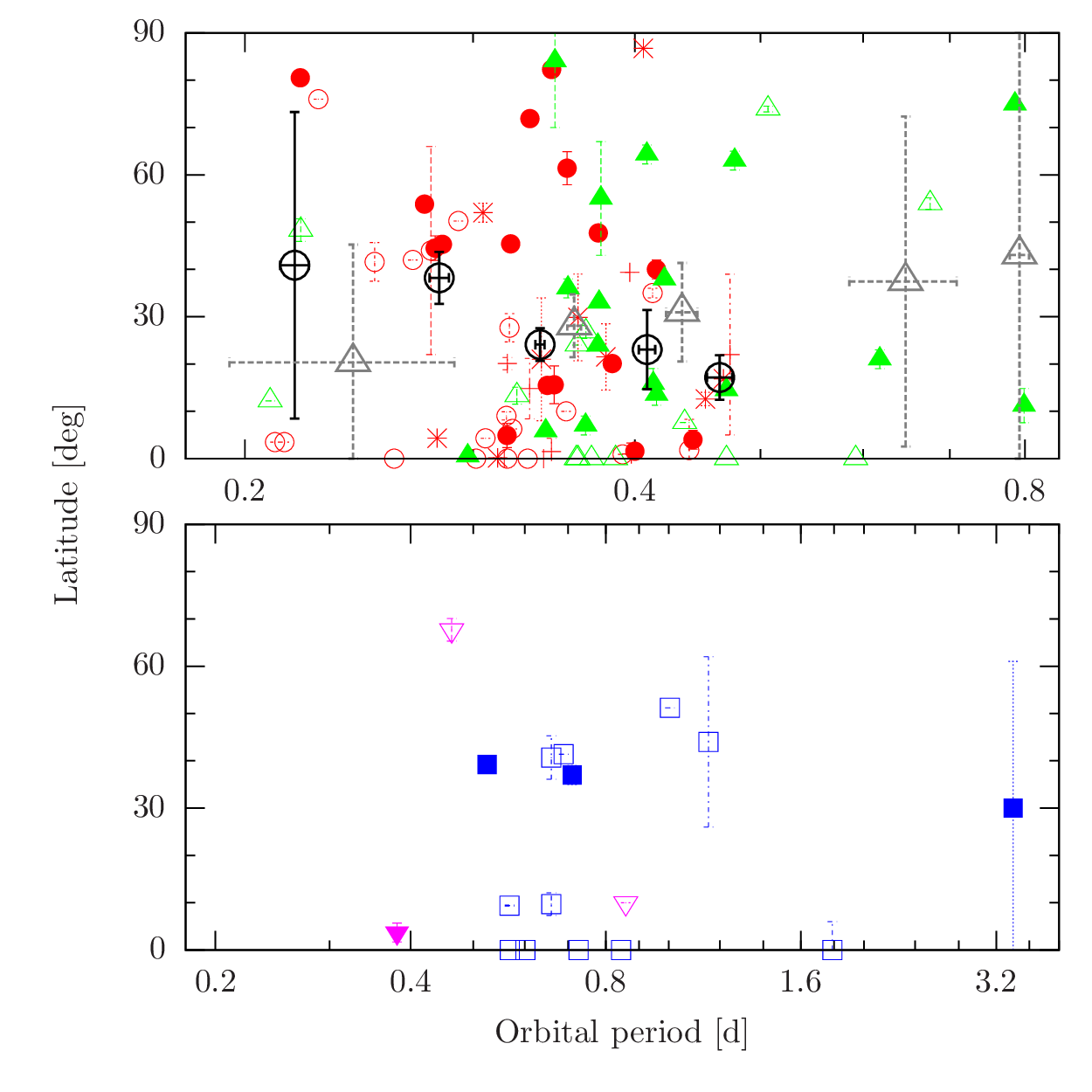}
\end{center}
\caption{
Orbital period versus latitude of cool spot. 
Symbols are the same as in figures \ref{Lon-Rad-Lat-CS} and \ref{Lat-Rad-CS}. 
}
\label{P-Lat-CS}
\end{figure}

\citet{Schussler1992-AA} concluded that magnetically active stars with rapid rotation exhibit magnetic flux eruption at high latitudes and polar spots. 
\citet{Schussler1996-AA} demonstrated that for slowly rotating stars flux emerges at lower latitudes and that the mean latitude of emergence shifts to higher latitudes for increasing stellar rotation rates. 
In figure \ref{P-Lat-CS}, we illustrate the scatter plot between the orbital period and the spot latitude. 
The orbital periods of contact binaries synchronize with their rotation periods 
because these binaries should be tidally locked. 
Hence, the rotation becomes rapid as the orbital period decreases. 
Figure \ref{P-Lat-CS} indicates that W-type sample binaries with a shorter orbital-period (i.e., fast rotators) tend to have cool spots at higher latitudes. 
The correlation coefficients for the W-type sample are 
$r_\textnormal{\scriptsize p}=-$0.219 with $p=0.134$ and 
$r_\textnormal{\scriptsize s}=-$0.172 with $p=0.243$. 
W-type sample binaries with spectroscopic mass-ratios show a stronger correlation: 
$r_\textnormal{\scriptsize p}=-$0.318 with $p=0.114$ and 
$r_\textnormal{\scriptsize s}=-$0.265 with $p=0.191$. 
We computed these coefficients on the basis of the W-type sample in which the binaries with fixed spot-parameters were excluded. 
This negative correlation agrees with the results from \citet{Schussler1992-AA} and \citet{Schussler1996-AA}, 
which can be explained by the dynamo theory. 
The A-type and semi-detached samples seem to have no clear correlation.

\begin{figure*}[tbp]
\begin{center}
\includegraphics[width=55mm]{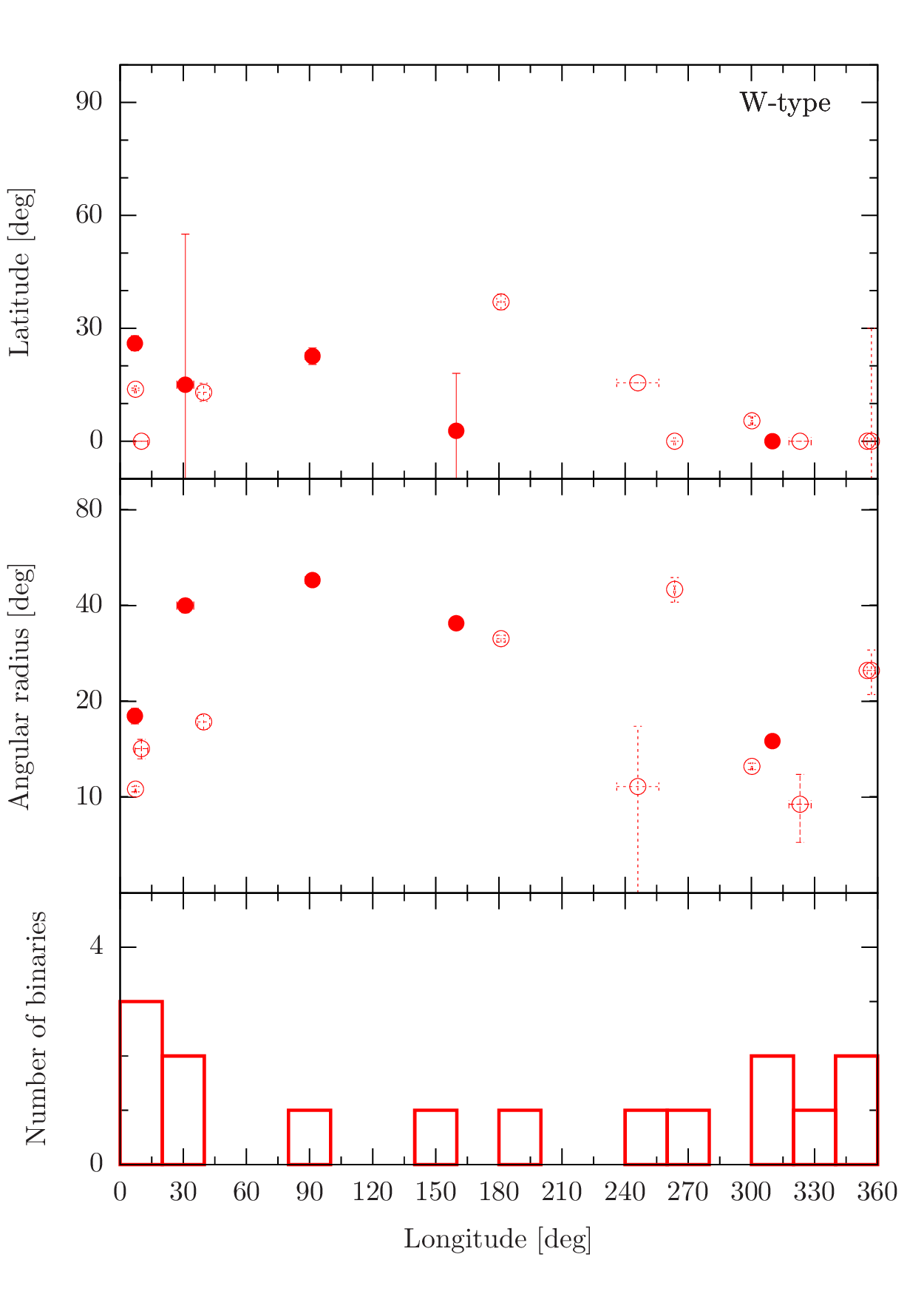}
\includegraphics[width=55mm]{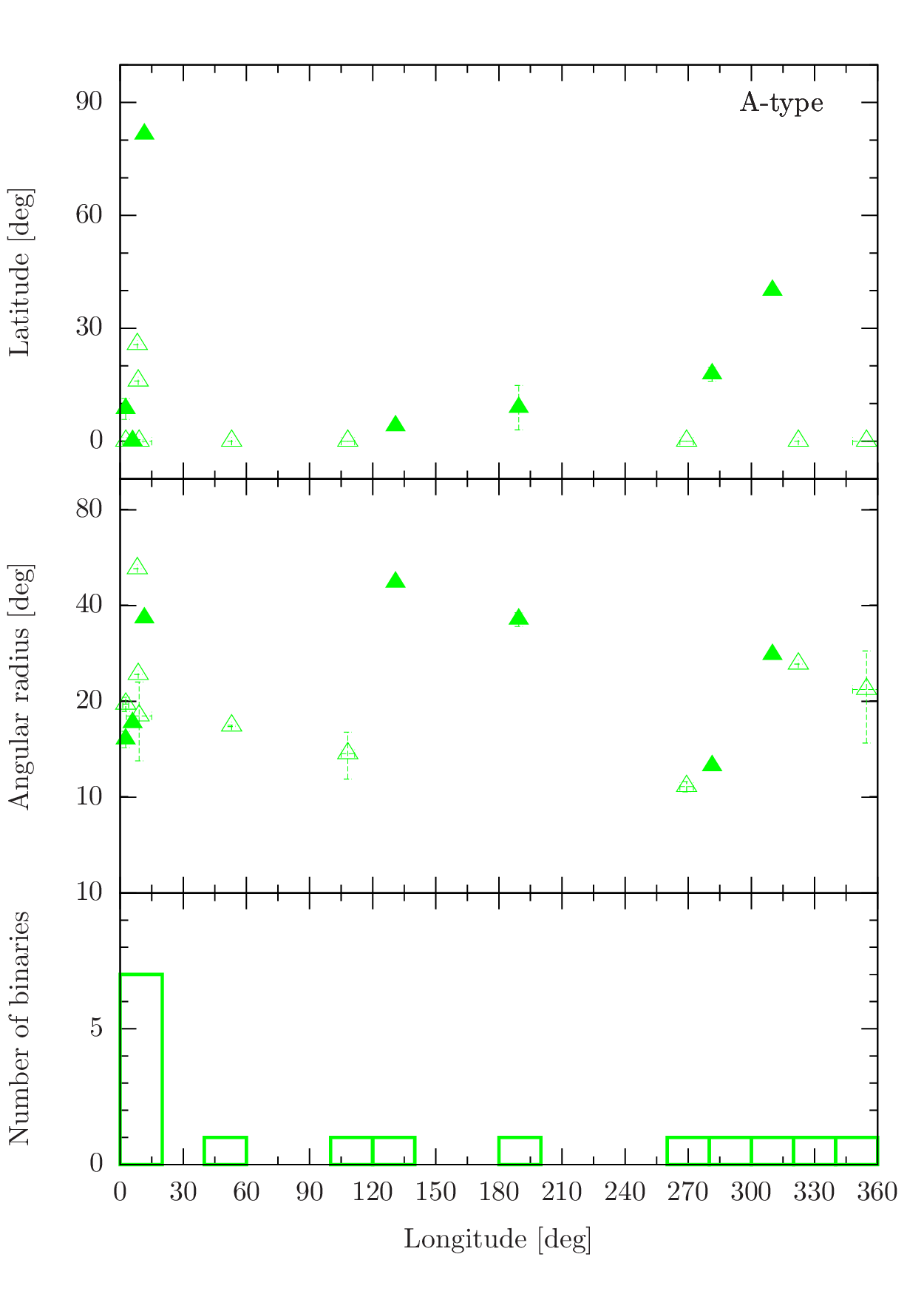}
\includegraphics[width=55mm]{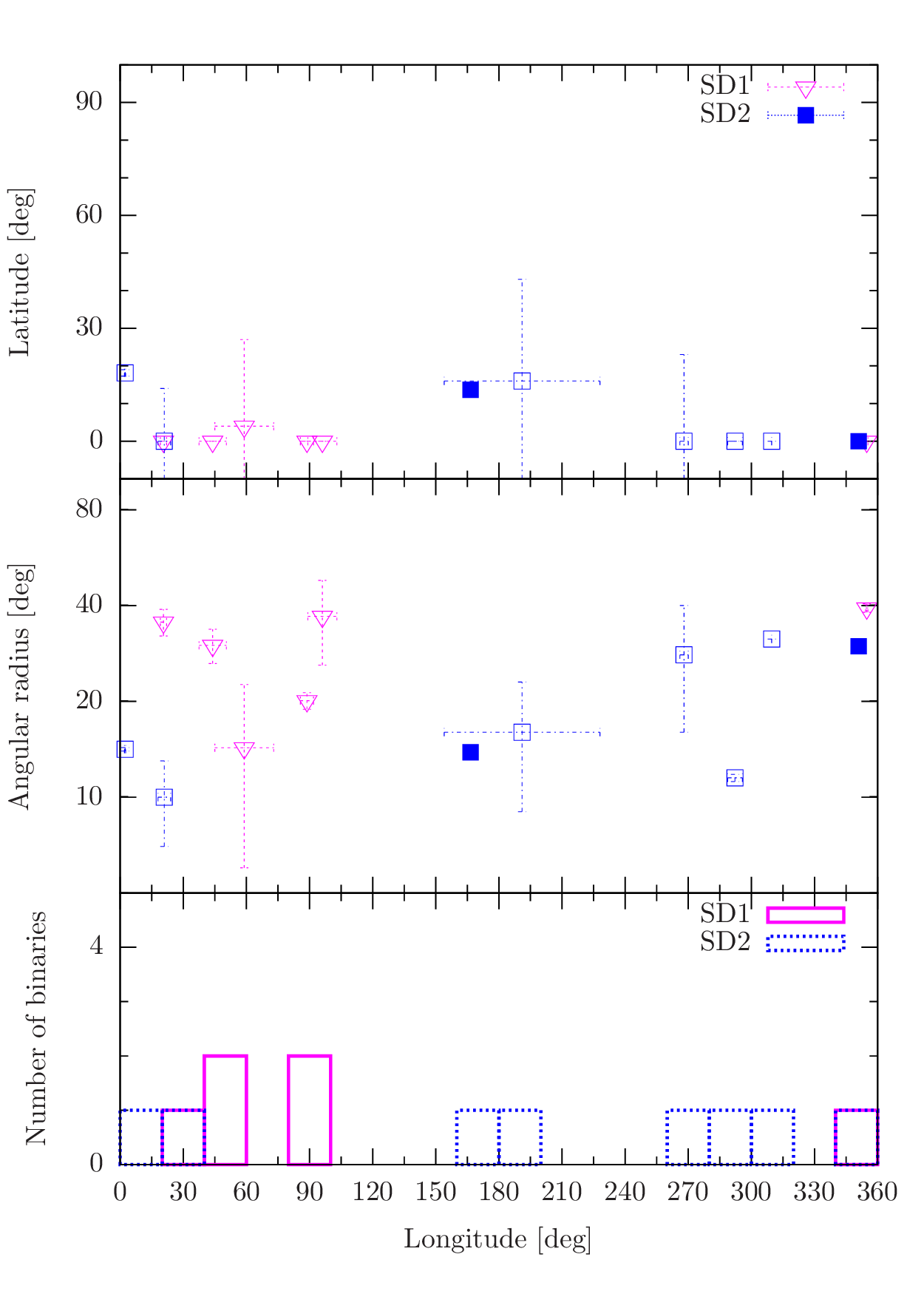}
\end{center}
\caption{
Longitudinal distributions of hot spots in contact and semi-detached binaries: 
hot spot distribution on the stellar surfaces (top panel), 
angular radius as a function of spot longitude (middle panel), and 
histograms of hot spot longitude (bottom panel). 
Symbols are the same as in figure \ref{Lon-Rad-Lat-CS}. 
}
\label{Lon-Rad-Lat-HS}
\end{figure*}
\subsection{Hot spot}
Most authors in table \ref{HS-binaries} concluded that hot spots were formed by mass-exchange between the two components of binary systems. 
However, the hot spots in at least five binary systems, namely AR Boo, EQ Tau, HL Aur, HR Boo, and LP Cep, 
were deduced to be formed by magnetic origin. 
\subsubsection{Position}
Figure \ref{Lon-Rad-Lat-HS} shows the scatter plots of longitude versus latitude and
longitude versus spot size, together with histograms of the number of binaries plotted as a function of longitude. 
The hot spots in both W- and A-type sample binaries are concentrated at $\phi\sim0^\circ$ and 
their latitudes are lower than $30^\circ$, 
where each spotted surface faces another component star. 
This tendency is reasonable when mass-exchange between two components of a binary system occurs through the first lagrange point. 
Also, SD1 sample binaries tend to have hot spots in the range $-90^\circ < \phi < 90^\circ$, with $\lambda<30^\circ$. 
However, the hot spots in SD2 binaries are widely distributed in longitude, 
which differs from the other samples. 

The hot spots of the five binaries mentioned in the above, which were deduced to be formed by magnetic origin, are located at the range $180^\circ<\phi<270^\circ$ except for AR Boo ($\phi=39^\circ_\cdot7$).
Thus, some of the other spots with a longitude between $90^\circ$ and $270^\circ$ may also be generated by magnetic activity. 

Only BX Dra has a polar spot with a latitude of $81^\circ_\cdot5$ despite the fact that all other binaries have hot spots with latitudes smaller than $30^\circ$. 
\citet{Park2013-PASJ}, nevertheless, surmised that the hot spot was caused by mass-transfer between the components rather than magnetic activity. 
Based on an $O-C$ diagram, they also confirmed that mass is transferred from the secondary to the primary at a rate of $2.74\times10^{-7} M_\odot$ yr$^{-1}$. 
If their claims are correct, this high-latitude spot is quite strange in our samples. 

The SD1 sample includes hot spots generally larger than those of the SD2 sample, 
although no such difference exists between the W- and A-type samples. 
No other clear association and tendency can be found in the relation between the longitude and the spot size.

\begin{figure}[tbp]
\begin{center}
\includegraphics[width=80mm]{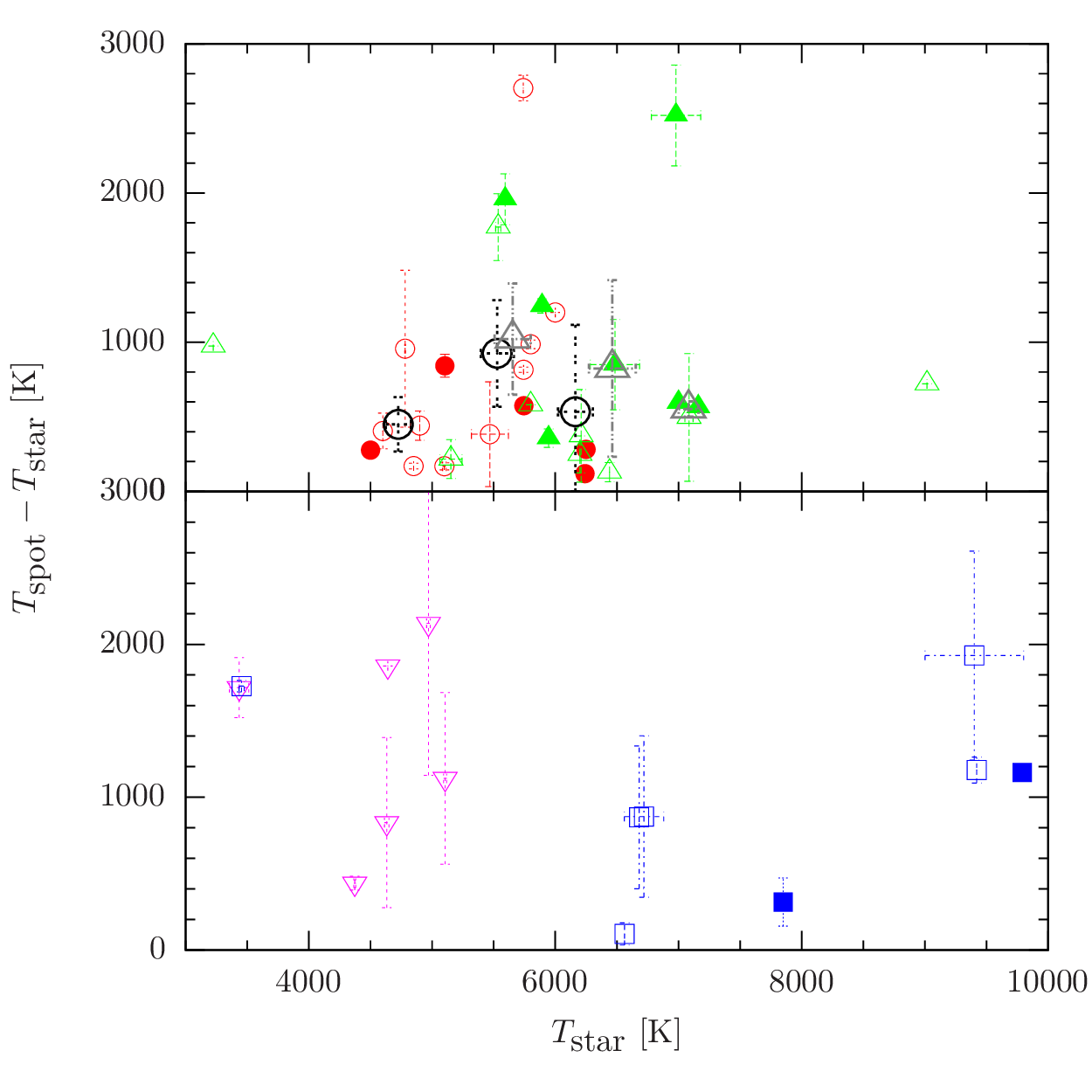}
\end{center}
\caption{
Photosphere temperature of spotted star versus temperature difference between hot spot and photosphere. 
Symbols are the same as in figures \ref{Lon-Rad-Lat-CS} and \ref{Lat-Rad-CS}. 
}
\label{Tstar-DeltaT-HS}
\end{figure}
\subsubsection{Temperature}
The relation between the temperature and the temperature difference is shown in figure \ref{Tstar-DeltaT-HS}. 
Although the temperature difference appears to increase with increasing temperature in the W-type sample, 
this is not significant 
($r_\textnormal{\scriptsize p}=$0.226 with $p=0.417$ and 
$r_\textnormal{\scriptsize s}=$0.157 with $p=0.576$). 
Spotted stars in the SD1 sample are generally cooler than those in the SD2 sample, unlike the case of cool spots.
This tendency arises from the fact that the SD1 (SD2) sample binaries have hot spots on the less-massive and cooler (the more-massive and hotter) components (see table \ref{Num-spotted}). 
In addition, SD1 sample binaries tend to have temperature differences larger than those of SD2 sample binaries.

\begin{figure}[tbp]
\begin{center}
\includegraphics[width=80mm]{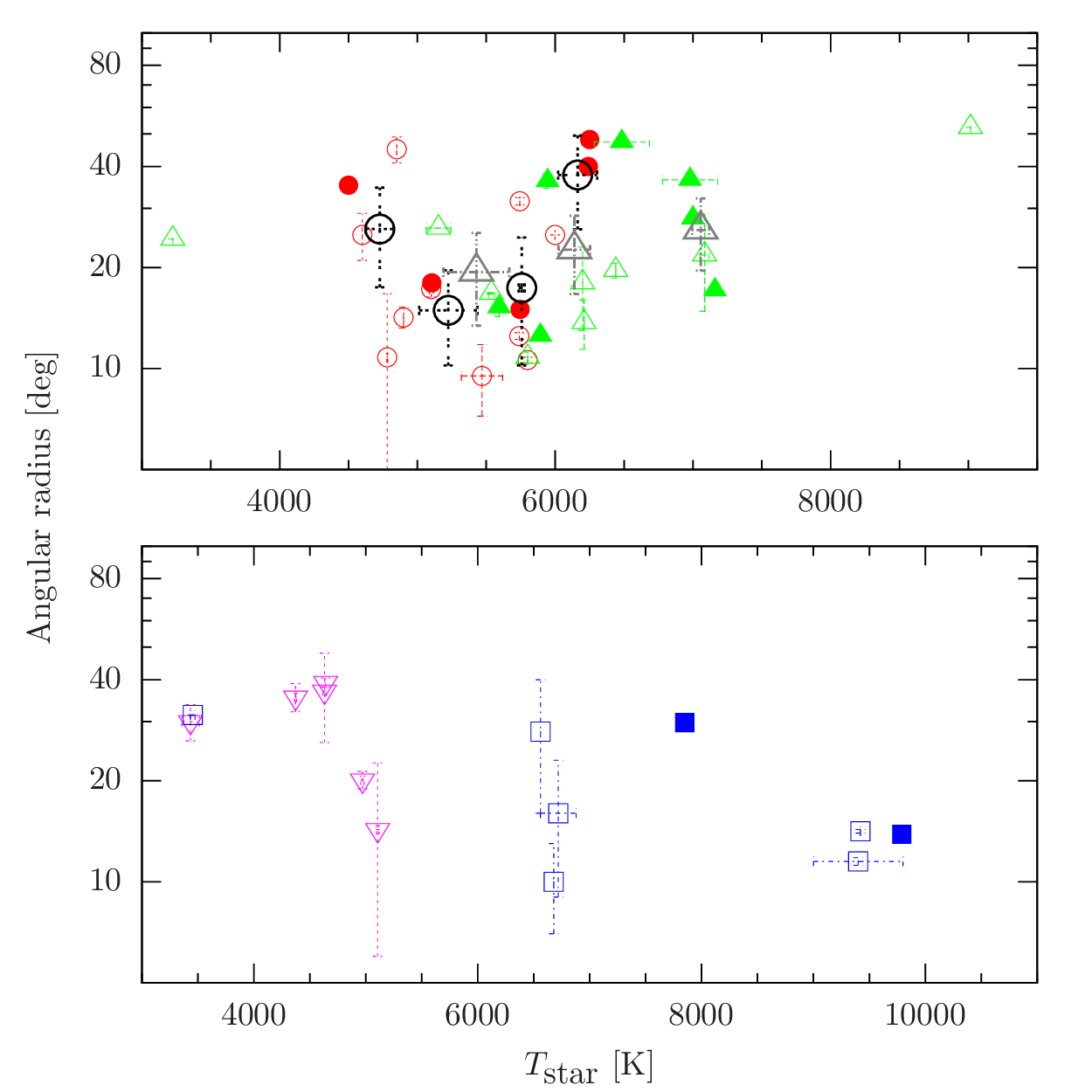}
\end{center}
\caption{
Photosphere temperature and angular radius of hot spot. 
Symbols are the same as in figures \ref{Lon-Rad-Lat-CS} and \ref{Lat-Rad-CS}. 
}
\label{Tstar-Rad-HS}
\end{figure}

In figure \ref{Tstar-Rad-HS}, we depict the scatter plot between the temperature and the the spot size. 
The W- and A-type samples are roughly separated by $T=5500$--$6000$ K and they seem to have different associations. 
In other words, the spot sizes of W- and A-type samples decreases and increases with increasing temperature respectively. 
The SD1 and SD2 samples also are separated by $T\sim6000$ K. 
No clear tendency is found in the semi-detached samples. 

Two W-type systems, namely AC Boo and TX Cnc, appear to have a tendency opposed to the general tendency of the W-type sample mentioned in the above. 
These systems have physical parameters of $T\sim6250$ K, $\Delta T\sim200$ K, and $\alpha \sim40^\circ$.  
Their positions in both $T$--$\Delta T$ and $T$--$\alpha$ scatter plots agree with those of the A-type binaries rather than those of the W-type binaries. 
\citet{Nelson2010-IBVS5951} concluded that AC Boo is an overcontact binary comprised of unevolved stars. 
In addition, TX Cnc is a member of the Praesepe open cluster, whose mean age is around 600 Myr \citep{Zhang2009-AJ}. 
Hence, because the two W-type systems comprise young stars, they may have the opposed tendency. 
When the two binaries are excluded from the W-type sample, 
although the positive correlation in the $T$--$\Delta T$ relation becomes significant 
($r_\textnormal{\scriptsize p}=$0.528 with $p=0.064$ and $r_\textnormal{\scriptsize s}=$0.588 with $p=0.035$), 
the negative correlation in the $T$--$\alpha$ relation is still insignificant 
($r_\textnormal{\scriptsize p}=-$0.323 with $p=0.281$ and $r_\textnormal{\scriptsize s}=-$0.300 with $p=0.320$).

\begin{figure}[tbp]
\begin{center}
\includegraphics[width=80mm]{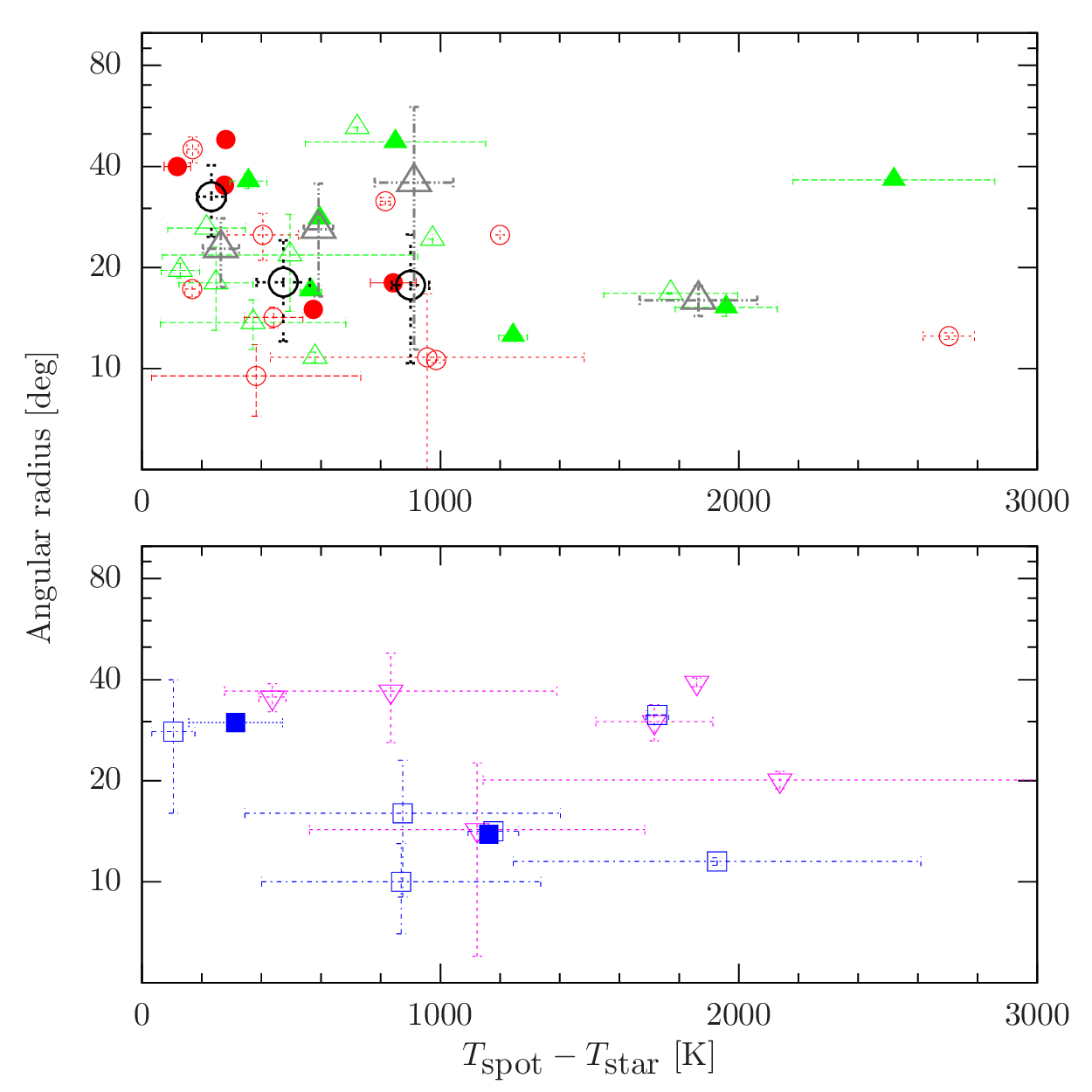}
\end{center}
\caption{
Temperature difference versus angular radius of hot spot. 
Symbols are the same as in figures \ref{Lon-Rad-Lat-CS} and \ref{Lat-Rad-CS}. 
}
\label{DeltaT-Rad-HS}
\end{figure}

A difference in correlation between the W- and A-type samples is also found in $\Delta T$--$\alpha$ relation (figure \ref{DeltaT-Rad-HS}). 
For the W-type sample, the angular radius decreases with increasing temperature difference 
($r_\textnormal{\scriptsize p}=-$0.445 with $p=0.096$ and 
$r_\textnormal{\scriptsize s}=-$0.531 with $p=0.042$). 
However, when the two systems that comprise young stars are excluded from the W-type sample, 
the negative correlation becomes weak 
($r_\textnormal{\scriptsize p}=-$0.356 with $p=0.233$ and 
$r_\textnormal{\scriptsize s}=-$0.366 with $p=0.219$). 
By contrast, the spot size for the A-type sample slightly increases as the temperature difference increases below $\Delta T=1000$ K 
($r_\textnormal{\scriptsize p}=$0.416 with $p=0.179$ and 
$r_\textnormal{\scriptsize s}=$0.329 with $p=0.297$). 
No strong correlation between the temperature difference and the spot size has been reported in several types of objects such as T Tauri stars and RS CVn-type stars (\cite{Bouvier1989-AA}; \cite{Strassmeier1992-ASPC}).

\begin{figure}[tbp]
\begin{center}
\includegraphics[width=80mm]{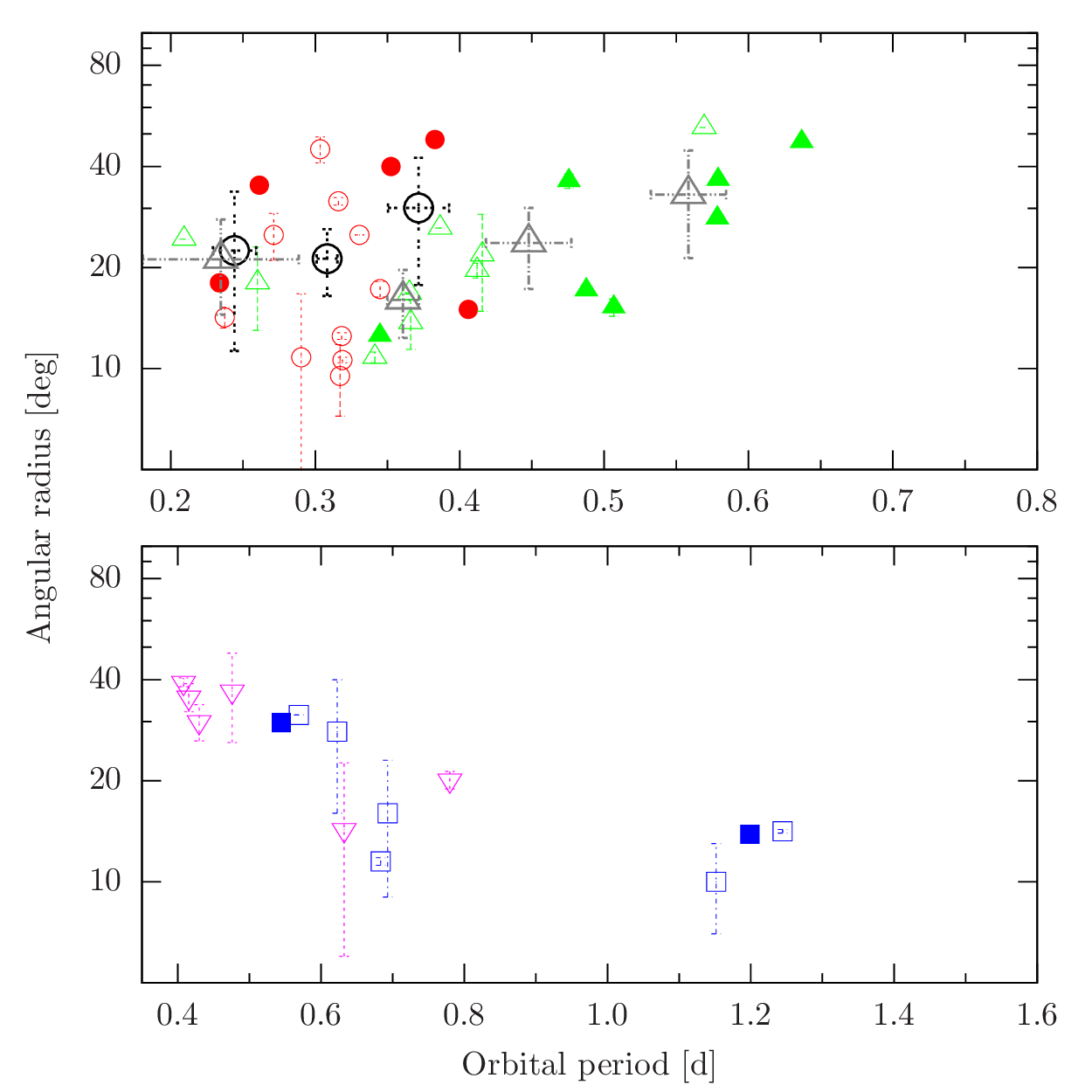}
\end{center}
\caption{
Orbital period versus angular radius of hot spot. 
Symbols are the same as in figures \ref{Lon-Rad-Lat-CS} and \ref{Lat-Rad-CS}. 
}
\label{P-Rad-HS}
\end{figure}
\subsubsection{Binary parameters}
Binary parameters are expected to be associated with hot-spot parameters, 
when mass transfer between component stars forms hot spots. 
This subsection examines correlations between the binary and hot-spot parameters. 

In figure \ref{P-Rad-HS}, we plot the orbital period versus the spot size. 
The A-type sample has a clear positive correlation above $P=0.3$--$0.4$ d 
($r_\textnormal{\scriptsize p}=$0.785 with $p<0.001$ and 
$r_\textnormal{\scriptsize s}=$0.798 with $p<0.001$). 
On the other hand, there is a negative correlation below the value 
($r_\textnormal{\scriptsize p}=-$0.984 with $p=0.016$ and 
$r_\textnormal{\scriptsize s}=-$0.800 with $p=0.200$). 
Although the W-type sample shows a weak positive correlation, this is not significant 
($r_\textnormal{\scriptsize p}=$0.200 with $p=0.476$ and 
$r_\textnormal{\scriptsize s}=$0.105 with $p=0.708$). 
Two sequences appear to exist in the distribution of the W-type sample. 
The sequence that have relatively small hot-spots ($\alpha<20^\circ$) overlaps with the distribution of the A-type sample. 
As discussed in section \ref{Discussion-HS}, the hot-spot origins of W-type binaries should be confused between mass transfer and magnetic activity, 
hence the two sequences may exist. 
The SD1 sample binaries tend to have orbital periods shorter than those of the SD2 sample binaries; and 
have a negative correlation 
($r_\textnormal{\scriptsize p}=-$0.822 with $p=0.045$ and 
$r_\textnormal{\scriptsize s}=-$0.771 with $p=0.072$).

\begin{figure}[tbp]
\begin{center}
\includegraphics[width=80mm]{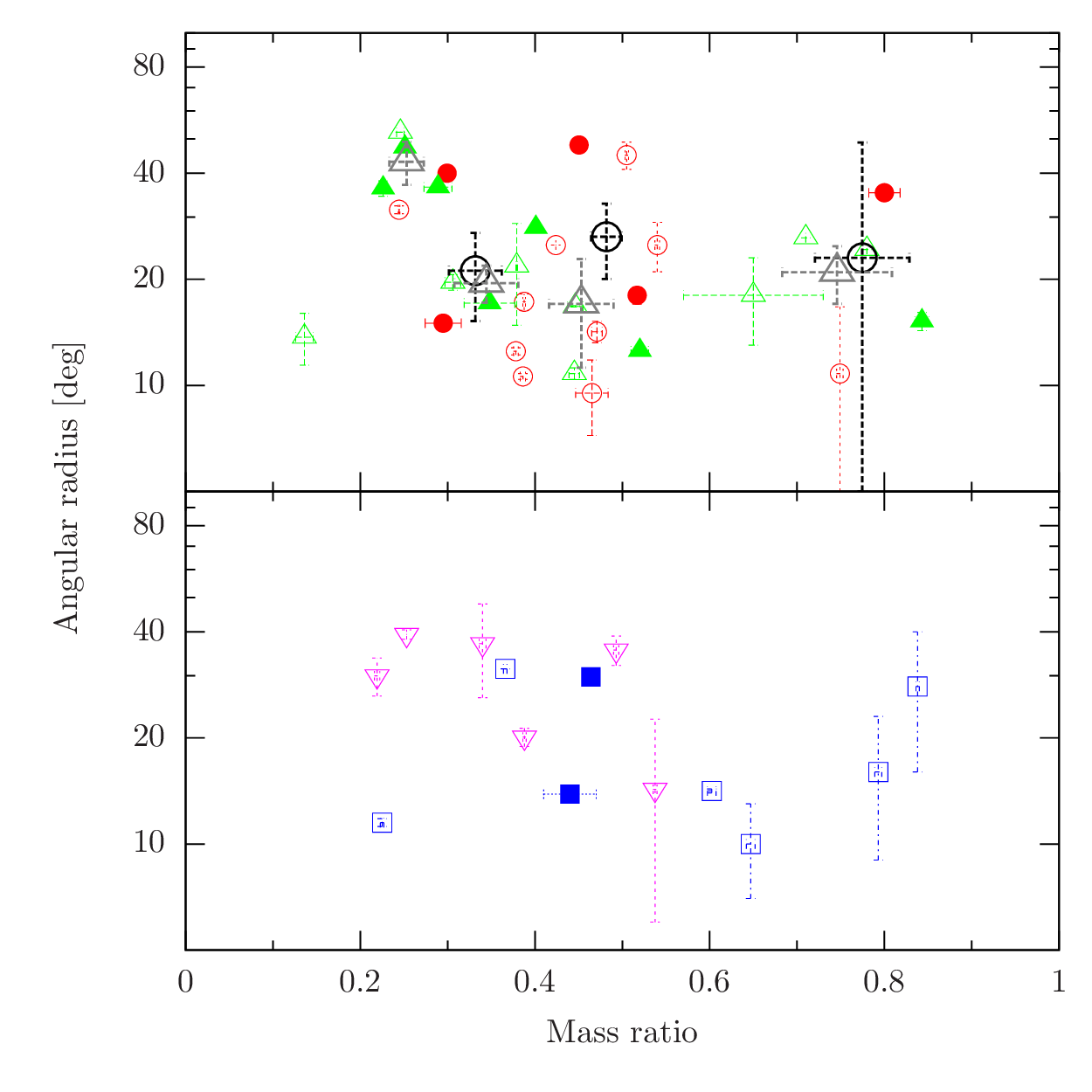}
\end{center}
\caption{
Mass ratio versus angular radius of hot spot. 
Symbols are the same as in figures \ref{Lon-Rad-Lat-CS} and \ref{Lat-Rad-CS}. 
}
\label{Mass_ratio-Rad-HS}
\end{figure}

Figure \ref{Mass_ratio-Rad-HS} shows the relation between the mass ratio and the spot size. 
We compute the mass ratio to fall between 0 and 1. 
The A-type sample has a negative correlation in the range $0.2<q<$0.6 
($r_\textnormal{\scriptsize p}=-$0.821 with $p=0.002$ and 
$r_\textnormal{\scriptsize s}=-$0.836 with $p=0.001$), 
whereas it has no correlation above $q=$0.6. 
The W-type sample shows no strong correlations. 
As for the semi-detached samples, the SD1 sample has a weak negative association 
($r_\textnormal{\scriptsize p}=-$0.534 with $p=0.275$ and 
$r_\textnormal{\scriptsize s}=-$0.543 with $p=0.266$).

\begin{figure}[tbp]
\begin{center}
\includegraphics[width=80mm]{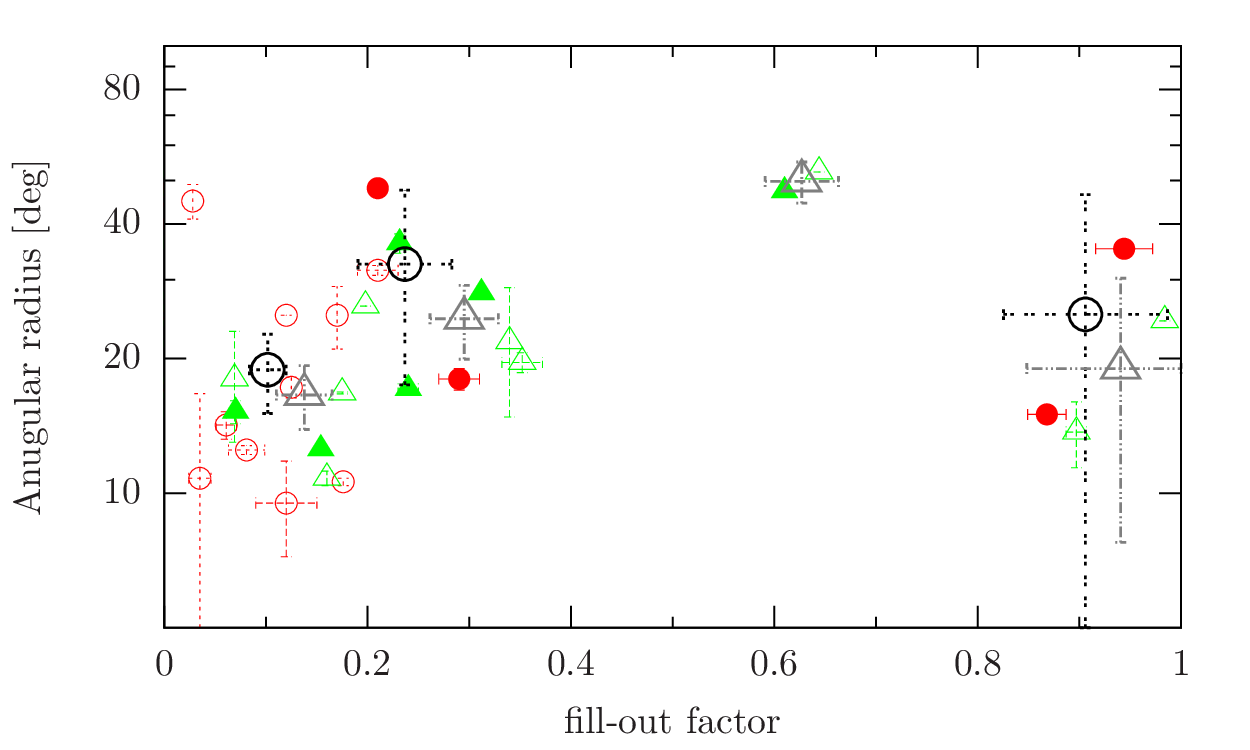}
\end{center}
\caption{
Fill-out factor versus angular radius of hot spot. 
Symbols are the same as in figures \ref{Lon-Rad-Lat-CS} and \ref{Lat-Rad-CS}. 
}
\label{F-Rad-HS}
\end{figure}

Another notable feature is in the relation between the fill-out factor and the spot size, which is illustrated in figure \ref{F-Rad-HS}. 
The W- and A-type samples have similar tendencies, i.e., 
the spot size increases with increasing fill-out factor. 
However, below $f=0.4$, the positive correlation of the W-type sample is weak 
($r_\textnormal{\scriptsize p}=0.142$ with $p=0.660$ and 
$r_\textnormal{\scriptsize s}=0.270$ with $p=0.396$), 
although that of the A-type sample is significant below $f=0.7$
($r_\textnormal{\scriptsize p}=$0.853 with $p<0.001$ and 
$r_\textnormal{\scriptsize s}=$0.747 with $p=0.003$). 
The weak correlation of the W-type sample is due to CE Leo which is the W-type sample binary with the smallest fill-out factor. 
CE Leo has a large hot-spot ($\alpha=45^\circ$), unlike the other systems with fill-out factors close to the smallest one. 
We surmise that magnetic activity, rather than mass transfer, formed the hot spot in CE Leo because of the following reasons. 
First, the spot is located at the opposite of substellar point and is indicative of a deep convective envelope due to its low temperature. 
Second, \citet{Kang2004-AJ} also concluded that CE Leo could have a cool or a hot spot; 
it was the result of chromospheric activity rather than a gas stream striking the surface of the mass-gaining component. 
Therefore, we deduce that magnetic activity formed the hot spot in CE Leo. 
If CE Leo is excluded from the sample, the correlation coefficients for the W-type sample are computed as 
$r_\textnormal{\scriptsize p}=0.492$ with $p=0.124$ and 
$r_\textnormal{\scriptsize s}=0.563$ with $p=0.071$ below $f=0.4$. 
These positive correlations are plausible when mass transfer generates hot spots. 

The association above $f=0.7$ may differ from that below the value. 
Four binaries with fill-out factors larger than $f=0.7$ are SS Ari, VZ Psc, DZ Psc, and 1SWASP J075102.16+342405.3. 
These binaries except DZ Psc are expected to have deep convective envelopes because their components have temperatures lower than 6000 K. 
The spotted component of DZ Psc has a temperature of 6210 K and also possibly have a convective envelope. 
Therefore, magnetic activity can generate the hot spots in the five binaries 
and the different association is due to the difference in spot formation mechanism. 
Alternatively, hot spot properties may differ between binaries with small and large fill-out factors.

\section{Discussion}\label{Discussion}
\subsection{Cool spot}
Our results indicate that the statistical properties of cool spots differ between W- and A-type contact binaries. 
In the $T$--$\Delta T$ and $T$--$\alpha$ relations, the associations of W-type sample binaries agree with those of late-type stars presented by \citet{Berdyugina2005-LRSP}. 
Section \ref{CS-Position} demonstrates that W-type binaries should have active longitudes separated by $180^\circ$;
active longitudes have been explained by the dynamo theory. 
The correlations in the $\theta$--$\alpha$ and $P$--$\lambda$ relations are also consistent with the properties that are derived from the dynamo theory. 
Because most of our W-type sample binaries have temperatures lower than 6000 K, 
it is natural that the dynamo theory explains the properties of cool spots in W-type binaries. 
Consequently, stellar dynamos are expected to form the cool spots in W-type binaries. 
By contrast, the cool spots of A-type sample binaries show positive correlations in the $T$--$\alpha$ and $P$--$\alpha$ relations and 
a negative correlation in the $\Delta T$--$\alpha$ relation;
their correlations differ from those of W-type sample binaries. 
Additionally, most of A-type sample binaries have temperatures higher than 6000 K. 
Therefore, other mechanism, which differs from stellar dynamo, may forms the cool spots in A-type binaries. 
A different mechanism for the spot formation in A-type systems is also supported by \citet{Pribulla2011-AN}. 

SD2 sample binaries also generally have cool spots on the components with temperatures lower than 6000 K. 
Nevertheless, their associations considerably differ from those of the W-type sample. 
Low statistics may allow their intrinsic associations to be unclear. 
Alternatively, mechanisms for generating cool spots differ between SD2 and W-type systems.

\subsection{Hot spot}\label{Discussion-HS}
The hot spots in A-type and SD2 sample binaries are likely to be formed by mass transfer rather than magnetic activity 
because these spots were present on components with $T>6000$ K. 
Accordingly, the correlations in the $T$--$\alpha$, $P$--$\alpha$, $q$--$\alpha$ and $f$--$\alpha$ relations of the A-type sample 
are expected to reflect the properties of hot spots formed by mass transfer; 
unless other mechanisms largely contributed to the formation of hot spots. 
These correlations indicate that the spot size increases with decreasing mass ratio ($0.2<q<0.6$) and increasing the temperature of the spotted stars, the orbital period ($P>0.3$--$0.4$ d), 
and the fill-out factor ($f<0.7$). 
The spotted components of SD2 sample binaries also have temperatures higher than 6000 K and 
mass transfer should mainly form their spots. 
Nevertheless, in contrast to the A-type sample, the SD2 sample shows no strong associations in any relations. 
The small size of the SD2 sample may make the intrinsic associations unclear. 
Alternatively, the hot-spot properties of SD2 binaries may differ from those of A-type binaries because of the difference in the configuration of binary systems. 

The hot spots in W-type and SD1 sample binaries can be formed by magnetic activity 
because these spots were present on components with $T<6000$ K. 
Accordingly, magnetic activity, as well as mass transfer, can form the hot spots in these systems. 
However, it is difficult to distinguish between the two mechanisms; 
our samples should include hot spots formed by both magnetic activity and mass transfer. 
This contamination contributes to dispersed distributions and weak associations. 
In practice, the associations found in the W-type sample are relatively weak. 
Consequently, these associations should be examined using a hot-spot sample in which spot formation mechanisms are well determined.

\subsection{The reliability of spot parameters}\label{Reliability}
Several combinations of spot parameters derived with light-curve modeling can produce similar light curves. 
The non-uniqueness problem has been discussed by previous studies (e.g. \cite{Eker1996-ApJ}). 
However, this does not indicate that light-curve modeling is unable to yield a unique solution. 
\citet{Eker1999-NewA} claimed that the empirical evidences implying non-uniqueness of spot solutions are caused by insufficient accuracy of observational data. 
\citet{Eker1999-ApJ} estimated that $\pm0.0001$ mag or better accuracy is required for deriving parameters with a reasonable accuracy. 
Because the accuracy of light curve data in this study is worse than the value, 
the uncertainty of spot parameters determined with light-curve modeling should be large. 
Nevertheless, our spotted-binary samples show several correlations and the correlations of W-type systems are consistent with the spot properties derived from the dynamo theory. 
In addition, these correlations differ at least between W- and A-type systems. 
If the parameters never reflect the intrinsic properties of binary systems, such correlations or differences should not be discovered. 
This indicates that the statistical analysis reduces the uncertainty of each parameter of individual system and 
their statistical correlations reasonably reflect the intrinsic properties of spotted binaries. 
However, our results should be verified with spot parameters which determined by other technique such as the Doppler imaging.

\section{Summary and Conclusions}\label{Conclusion}
We have investigated the statistical properties of starspots in eclipsing binaries on the basis of parameters determined by synthetic light-curve analysis. 
Our results indicate that the magnetic activity caused by stellar dynamos should form the cool spots in W-type contact binaries. 
By contrast, the A-type and semi-detached samples show associations differing from those of the W-type sample. 
These different associations indicate that the mechanism for forming the cool spots in A-type and semi-detached binaries differs from that in W-type binaries. 

Hot spot properties also differ between the W- and A-type samples. 
We found clear correlations in the $T$--$\alpha$, $P$--$\alpha$, $q$--$\alpha$, and $f$--$\alpha$ relations for the A-type sample and in the $T$--$\Delta T$, $\Delta T$--$\alpha$, and $f$--$\alpha$ relations for the W-type sample. 
We infer that magnetic activity, as well as mass transfer, can contribute to the formation of the hot spots in W-type binaries unlike the case of A-type binaries. 
The hot spot properties of SD1 binaries also differ from those of SD2 binaries. 
SD1 binaries seems to have spot activity stronger than SD2 binaries 
because the mean spot-size and the temperature difference between spot and photosphere are larger in the SD1 sample than in the SD2 sample. 

Both cool and hot spot properties of the SD1 and SD2 samples seem to be different. 
However, the semi-detached samples suffer from low statistics. 
Accordingly, further investigation with a large-size sample is required. 

A problem of light curve modeling is that its solution is not usually unique. 
This allows starspot parameters to have alternative ones and the uncertainties of parameters tend to be large. 
Statistical analysis reduces the effect of the uncertainty of each parameter. 
In practice, several associations found in the W-type sample are consistent with each other. 
This suggests that a statistical analysis with spot parameters determined by light curve modelling is effective for investigating the general trends of starspots. 
Nevertheless, uncertainties of the statistical analysis might still remain large, particularly in the case of low statistics. 
Hence, the starspot properties in this paper should be further examined with a sample of spotted binaries whose parameters are determined by other method such as the Doppler imaging technique.

{
\begin{longtable}{lccrrrrrccrrrrr}
\caption{
Parameters of binary systems with cool spots.  \label{CS-binaries}}
  \hline
\multicolumn{1}{c}{Object\footnotemark[$*$]} & \multicolumn{1}{c}{Tp\footnotemark[$\dag$]} & \multicolumn{1}{c}{STp\footnotemark[$\ddag$]}	& \multicolumn{1}{c}{$P$\footnotemark[$\S$]}	& \multicolumn{1}{c}{$q$\footnotemark[$\|$]} & \multicolumn{1}{c}{$f$\footnotemark[$\#$]} & \multicolumn{1}{c}{$T_1$\footnotemark[$*$$*$]} & \multicolumn{1}{c}{$T_2$\footnotemark[$\dag\dag$]} & \multicolumn{1}{c}{L/M\footnotemark[$\ddag\ddag$]} & \multicolumn{1}{c}{C/H\footnotemark[$\S\S$]} & \multicolumn{1}{c}{$\alpha$\footnotemark[$\|\|$]} 	 & \multicolumn{1}{c}{$T_\textnormal{\scriptsize spot}$\footnotemark[$\#\#$]} & \multicolumn{1}{c}{lon} & \multicolumn{1}{c}{colat} & \multicolumn{1}{c}{Ref\footnotemark[$***$]} \\
		 &		& 			& \multicolumn{1}{c}{(d)}	& 	  &	   	& \multicolumn{1}{c}{(K)}	& \multicolumn{1}{c}{(K)}	& 	  & 	& \multicolumn{1}{c}{(deg)}	 & \multicolumn{1}{c}{(K)}								 & \multicolumn{1}{c}{(deg)} 	 & \multicolumn{1}{c}{(deg)}	  & 		   \\
\hline
\endfirsthead
\hline
\multicolumn{1}{c}{Object} & \multicolumn{1}{c}{Tp} & \multicolumn{1}{c}{STp}	& \multicolumn{1}{c}{$P$}	& \multicolumn{1}{c}{$q$} & \multicolumn{1}{c}{$f$} & \multicolumn{1}{c}{$T_1$} & \multicolumn{1}{c}{$T_2$} & \multicolumn{1}{c}{L/M} & \multicolumn{1}{c}{C/H} & \multicolumn{1}{c}{$\alpha$} 	 & \multicolumn{1}{c}{$T_\textnormal{\scriptsize spot}$} & \multicolumn{1}{c}{lon} & \multicolumn{1}{c}{colat} & \multicolumn{1}{c}{Ref} \\
		 &		& 			& \multicolumn{1}{c}{(d)}	& 	  &	   	& \multicolumn{1}{c}{(K)}	& \multicolumn{1}{c}{(K)}	& 	  & 	& \multicolumn{1}{c}{(deg)}	 & \multicolumn{1}{c}{(K)}								 & \multicolumn{1}{c}{(deg)} 	 & \multicolumn{1}{c}{(deg)}	  & 		   \\
\hline
\endhead
\hline
\endfoot
\hline
\multicolumn{15}{p{8cm}}{\parbox{170mm}{
\footnotesize
\footnotemark[$*$]Full names of abbreviation: 
GSC 1537-1557 (GSC 1537), 
GSC 3551-1708 (GSC 3551), 
GSC 03526-01995 (GSC 03526), 
NSVS 1461538 (NSVS 146), 
UCAC4 436-062932 (UCAC4 436), 
1SWASP J015100.23-100524.2 (SW J015), 
1SWASP J074658.62+224448.5 (SW J074), 
1SWASP J075102.16+342405.3 (SW J075), 
2MASS 02272637+1156494 (2MASS 022). 
}}\\ 
\multicolumn{15}{p{8cm}}{\parbox{170mm}{
\footnotesize\footnotemark[$\dag$]Types introduced by \citet{Kopal1955-AnAp}. 
Contact and semi-detached binaries are denoted by C and SD, respectively. 
The symbol of SD1 (SD2) indicates that the more (less) massive compoent filling its Roche lobe. 
}}\\
\multicolumn{15}{p{8cm}}{\footnotesize\footnotemark[$\ddag$]Subtypes for contact binaries, which are based on \citet{Binnendijk1970-VA}. }\\
\multicolumn{15}{l}{\footnotesize\footnotemark[$\S$]Orbital period. }\\
\multicolumn{15}{p{8cm}}{\parbox{170mm}{\footnotesize\footnotemark[$\|$]Mass ratio. The symbol "s" refers to a value which is determined with spectroscopic data.}}\\
\multicolumn{15}{l}{\footnotesize\footnotemark[$\#$]Fill-out factor. }\\
\multicolumn{15}{p{8cm}}{\parbox{170mm}{\footnotesize\footnotemark[$**$]Temperature of spotted star. The symbol "f" refers to a value which was fixed during light-curve modeling.}}\\
\multicolumn{15}{l}{\footnotesize\footnotemark[$\dag\dag$]Temperature of unspotted star. }\\
\multicolumn{15}{p{8cm}}{\parbox{170mm}{\footnotesize\footnotemark[$\ddag\ddag$]The symbol L (M) stands for the less (more) massive component having a starspot. }}\\
\multicolumn{15}{p{8cm}}{\parbox{170mm}{\footnotesize\footnotemark[$\S\S$]The symbol C (H) stands for the cooler (hotter) component having a starspot. }}\\
\multicolumn{15}{l}{\footnotesize\footnotemark[$\|\|$]Angular radius of spot. }\\
\multicolumn{15}{l}{\footnotesize\footnotemark[$\#\#$]Spot temperature. }\\
\multicolumn{15}{p{8cm}}{\parbox{170mm}{
\footnotesize
\footnotemark[$***$]References: 
1. \citet{Lee2011-PASP}; 2. \citet{Barone1993-ApJ}; 
3. \citet{Djurasevic2000-AA}; 4. \citet{Hrivnak1988-ApJ}; 
5. \citet{Zhang2005-AJ};  
6. \citet{Caliskan2014-AJ}; 7. \citet{Pribulla2006-AJ}; 
8. \citet{Lee2009-AJ};  
9. \citet{Alton2018-MNRAS};  
10. \citet{Gurol2005-NewA}; 11. \citet{Hrivnak1993-ASPC}; 
12. \citet{Broens2013-MNRAS};  
13. \citet{Niarchos1997-AAS};  
14. \citet{Kim2004-PASP};  
15. \citet{Kalomeni2007-AJ}; 
16. \citet{Torres2003-AJ}; 
17. \citet{Djurasevic2011-AA}; 18. \citet{Rucinski2008-AJ}; 
19. \citet{Qian2008-AJ2493}; 20. \citet{Lu1988-AcApS}; 
21. \citet{Qian2013-ApJS}; 
22. \citet{Yang2008-AJ}; 
23. \citet{Wang2018-PASJ}; 
24. \citet{Lee2004-MNRAS}; 25. \citet{Samec1991-AJ}; 
26. \citet{Zola2010-MNRAS}; 27. \citet{Rucinski1977-PASP}; 
28. \citet{Gazeas2006-AcA}; 29. \citet{Rucinski1999-AJ}; 
30. \citet{VanHamme2001-AJ}; 
31. \citet{Wang2014-AJ}; 
32. \citet{Niarchos1992-AA}; 
33. \citet{Lee2015-AJ}; 
34. \citet{Manimanis2002-ApSS}; 35. \citet{Zasche2009-OEJV};  
36. \citet{Djurasevic2013-AJ}; 
37. \citet{Djurasevic2006-PASA}; 38. \citet{Rucinski2002-AJ}; 
39. \citet{Yang2011-RAA}; 
40. \citet{Yang2002-AJ}; 41. \citet{Rucinski2001-AJ}; 
42. \citet{Lu2001-AJ}; 
43. \citet{Siwak2010-AcA}; 
44. \citet{Liu2012-PASJ};  
45. \citet{Gazeas2005-AcA}; 
46. \citet{Wang2015-AJ};  
47. \citet{Xiang2015-AJ9};  
48. \citet{Nelson2002-IBVS};  
49. \citet{Liao2012-AJ};  
50. \citet{Yang2017-PASJ};  
51. \citet{Lee2010-AJ898};  
52. \citet{Gurol2016-NewA46-31}; 
53. \citet{Lee2009-PASP}; 
54. \citet{Gurol2015-NewA}; 
55. \citet{Rucinski2005-AJ}; 
56. \citet{Gazeas2007-ASPC}; 57. \citet{Rucinski2000-AJ}; 
58. \citet{Zhu2005-AJ}; 
59. \citet{Kim2005-PASP}; 
60. \citet{Erdem2014-NewA}; 
61. \citet{Kim2016-JASS}; 
62. \citet{Djurasevic1998-ApSS}; 63. \citet{Hrivnak1989-ApJ}; 
64. \citet{Michaels2017-JAVSO133}; 
65. \citet{Lee2008-PASP}; 
66. \citet{Liu2014-ASPC}; 67. \citet{Zasche2014-AJ}; 
68. \citet{Ulas2009-ApSS}; 
69. \citet{Yang2003-AJ}; 
70. \citet{Kim2003-AJ}; 
71. \citet{Albayrak2004-AA}; 72. \citet{Zhai1989-ChAA}; 
73. \citet{Lee2007-PASP}; 
74. \citet{Milone1991-ApJ}; 
75. \citet{Li2015-AJ}; 
76. \citet{Christopoulou2011-AJ}; 77. \citet{Pribulla2009-AJ137-3646}; 
78. \citet{Djurasevic2001-AA}; 79. \citet{Zhai1988-ApSS}; 
80. \citet{Djurasevic2016-AJ}; 
81. \citet{Manzoori2015-ApSS}; 
82. \citet{Ekmekci2012-NewA}; 
83. \citet{Liu2015-AJ};  
84. \citet{Ekmekci2012-NewA}; 
85. \citet{Samec2017-JAVSO};  
86. \citet{Gurol2016-NewA47-57}; 
87. \citet{Rucinski2003-AJ}; 
88. \citet{Pribulla2007-AJ}; 
89. \citet{Jeong2013-JASS}; 
90. \citet{Djurasevic2008-RMxAA}; 
91. \citet{Kang2001-MNRAS}; 
92. \citet{Zhang2015-AJ}; 
93. \citet{Yang2001-AJ}; 
94. \citet{Lu1999-AJ}; 
95. \citet{Xiao2016-PASJ}; 96. \citet{Pych2004-AJ}; 
97. \citet{Hu2018-NewA}; 
98. \citet{Kallrath2006-AA}; 
99. \citet{Li2018-PASP}; 
100. \citet{Liu2011-AJ}; 
101. \citet{Kaszas1998-AA}; 
102. \citet{Virnina2012-OAP};  
103. \citet{Barden1987-ApJ}; 
104. \citet{Qian2011-AJ}; 
105. \citet{Michaels2017-JAVSO}; 
106. \citet{Yang1999-AAS}; 
107. \citet{Oh2006-MNRAS}; 
108. \citet{Qian2015-AJ}; 
109. \citet{Jiang2015-AJ}; 
110. \citet{Jiang2015-PASJ}; 
111. \citet{Liu2015-AJ111}; 
}}
\endlastfoot
                      AA UMa &   C &   W & $ 0.46813$ & $^\textnormal{s} 1.82$ & $ 0.14$ & $ 5920$ & $ 5964$ &  M  &  C  & $  20.6$ & $ 5195$ & $  98.2$ & $                 73.1$ &  1, 2 \\ 
                      AB And &   C &   W & $ 0.33189$ & $^\textnormal{s} 0.49$ & $ 0.15$ & $ 5450$ & $ 5705$ &  M  &  C  & $  32.3$ & $ 3815$ & $ 244.4$ & $                 18.1$ &  3, 4 \\ 
                      AH Cnc &   C &   A & $ 0.36046$ & $                0.15$ & $ 0.65$ & $ 6300$ & $ 6354$ &  M  &  C  & $   8.8$ & $ 4788$ & $  79.8$ & $^\textnormal{f}  90.0$ &  5 \\ 
                      AK Her &   C &   A & $ 0.42152$ & $^\textnormal{s} 0.28$ & $ 0.33$ & $ 6500$ & $ 6180$ &  M  &  H  & $  22.1$ & $ 5246$ & $ 305.4$ & $                127.9$ &  6, 7 \\ 
                      AR Boo &   C &   W & $ 0.34487$ & $                2.58$ & $ 0.12$ & $ 5100$ & $ 5382$ &  M  &  C  & $  14.4$ & $ 4718$ & $ 160.4$ & $                 91.5$ &  8 \\ 
                      AR CrB &   C &   W & $ 0.39735$ & $                0.84$ &   ---   & $ 5640$ & $ 5690$ &  M  &  C  & $  13.5$ & $ 4912$ & $ 145.1$ & $                 91.0$ &  9 \\ 
                      AU Ser &   C &   A & $ 0.38650$ & $^\textnormal{s} 0.71$ & $ 0.20$ & $ 5153$ & $ 5495$ &  L  &  C  & $  25.0$ & $ 3684$ & $ 241.9$ & $^\textnormal{f}  90.0$ & 10, 11 \\ 
                      AW CrB &   C &   A & $ 0.36094$ & $                0.10$ & $ 0.75$ & $ 6700$ & $ 6808$ &  M  &  C  & $  21.0$ & $ 6432$ & $ 290.0$ & $                114.0$ & 12 \\ 
                      AW Vir &   C &   W & $ 0.35400$ & $                1.31$ & $ 0.08$ & $ 5872$ & $ 6200$ &  M  &  C  & $  14.6$ & $ 4228$ & $ 309.5$ & $                 80.0$ & 13 \\ 
                      AX Dra & SD2 & --- & $ 0.56816$ & $                0.63$ &   ---   & $ 4951$ & $ 6850$ &  L  &  C  & $  19.0$ & $ 4258$ & $ 330.0$ & $                 80.6$ & 14 \\ 
                      BB Peg &   C &   W & $ 0.36150$ & $^\textnormal{s} 2.70$ & $ 0.34$ & $ 5955$ & $ 6250$ &  M  &  C  & $  14.3$ & $ 5479$ & $ 273.9$ & $                 60.2$ & 15 \\ 
           BD +05$^\circ$706 & SD2 & --- & $18.89880$ & $^\textnormal{s} 0.21$ &   ---   & $ 5000$ & $ 4640$ &  M  &  H  & $  33.8$ & $^\textnormal{f} 4350$ & $ 112.7$ & $^\textnormal{f}  45.0$ & 16 \\ 
           BD +73$^\circ$142 &   C &   W & $ 0.27520$ & $^\textnormal{s} 0.66$ & $ 0.10$ & $ 4640$ & $ 4900$ &  M  &  C  & $  33.3$ & $ 3248$ & $ 169.6$ & $                 36.2$ & 17, 18 \\ 
                      BI CVn &   C &   W & $ 0.38421$ & $^\textnormal{s} 0.41$ & $ 0.18$ & $ 6720$ & $ 6700$ &  L  &  H  & $  20.3$ & $ 4254$ & $ 186.5$ & $                 69.9$ & 19, 20 \\ 
                      BI Vul &   C &   W & $ 0.25183$ & $                1.04$ & $ 0.09$ & $ 4460$ & $ 4600$ &  M  &  C  & $  23.4$ & $ 3748$ & $  84.1$ & $                 48.4$ & 21 \\ 
                      BS Cas &   C &   W & $ 0.44047$ & $                0.28$ & $ 0.32$ & $ 5637$ & $ 6100$ &  M  &  C  & $  18.3$ & $ 4622$ & $ 260.7$ & $                 88.2$ & 22 \\ 
                      BU Vul & SD2 & --- & $ 0.56899$ & $                0.37$ &   ---   & $ 5940$ & $ 3454$ &  M  &  H  & $  28.9$ & $ 4158$ & $ 286.6$ & $                 90.0$ & 23 \\ 
                      BX Peg &   C &   W & $ 0.28042$ & $^\textnormal{s} 0.37$ & $ 0.23$ & $ 5300$ & $ 5536$ &  M  &  C  & $  18.0$ & $ 4447$ & $  66.9$ & $                 45.5$ & 24, 25 \\ 
                      CC Com &   C &   W & $ 0.22069$ & $^\textnormal{s} 0.52$ & $ 0.18$ & $ 4300$ & $ 4263$ &  M  &  H  & $  50.6$ & $ 3169$ & $ 130.7$ & $                170.5$ & 26, 27 \\ 
                      CK Boo &   C &   A & $ 0.35516$ & $^\textnormal{s} 0.11$ & $ 0.91$ & $ 6150$ & $ 6163$ &  M  &  C  & $  56.8$ & $ 6027$ & $  84.0$ & $                126.0$ & 28, 29 \\ 
                      CN And & SD1 & --- & $ 0.46279$ & $^\textnormal{s} 0.39$ &   ---   & $ 6500$ & $ 5922$ &  M  &  H  & $  32.9$ & $^\textnormal{f} 4225$ & $  16.9$ & $  22.3$ & 30 \\ 
                      CW Cas &   C &   W & $ 0.31886$ & $                2.06$ & $ 0.22$ & $ 4950$ & $ 5309$ &  M  &  C  & $  24.0$ & $ 3712$ & $ 236.2$ & $                 90.0$ & 31 \\ 
                      DF Hya &   C &   W & $ 0.33060$ & $                2.36$ & $ 0.12$ & $ 5851$ & $ 6000$ &  M  &  C  & $  10.0$ & $ 4096$ & $ 270.0$ & $^\textnormal{f}  90.0$ & 32 \\ 
                      DK Cyg &   C &   A & $ 0.47069$ & $^\textnormal{s} 0.33$ &   ---   & $ 7500$ & $ 7011$ &  M  &  H  & $  33.7$ & $ 7088$ & $ 182.0$ & $                 75.5$ & 29, 33 \\ 
                      DM Del & SD2 & --- & $ 0.84467$ & $                0.26$ &   ---   & $ 5117$ & $ 8770$ &  L  &  C  & $  22.3$ & $ 3326$ & $ 354.0$ & $^\textnormal{f}  90.0$ & 34, 35 \\ 
                      DU Boo &   C &   A & $ 1.05588$ & $^\textnormal{s} 0.23$ & $ 0.50$ & $ 7610$ & $ 7850$ &  L  &  C  & $  86.0$ & $ 5784$ & $ 349.0$ & $                 61.0$ &  7, 36 \\ 
                      EE Cet &   C &   W & $ 0.37992$ & $^\textnormal{s} 0.32$ & $ 0.33$ & $ 6095$ & $ 6314$ &  M  &  C  & $  15.8$ & $ 5546$ & $  58.2$ & $                 68.5$ & 37, 38 \\ 
                      EI CVn &   C &   W & $ 0.26077$ & $                0.46$ & $ 0.21$ & $ 4410$ & $ 4341$ &  M  &  H  & $  15.9$ & $ 3837$ & $ 239.5$ & $^\textnormal{f}  90.0$ & 39 \\ 
                      EQ Tau &   C &   A & $ 0.34135$ & $^\textnormal{s} 0.44$ & $ 0.19$ & $ 5735$ & $ 5800$ &  L  &  C  & $  18.6$ & $ 4588$ & $ 261.8$ & $                 95.8$ & 40, 41 \\ 
                      ET Leo &   C &   W & $ 0.34650$ & $^\textnormal{s} 0.34$ & $ 0.55$ & $ 5112$ & $ 5500$ &  M  &  C  & $  11.5$ & $ 3118$ & $  41.5$ & $                 74.4$ & 28, 38 \\ 
                      EX Leo &   C &   A & $ 0.40860$ & $^\textnormal{s} 0.20$ & $ 0.35$ & $ 6340$ & $ 6110$ &  M  &  H  & $  32.8$ & $ 3506$ & $  85.5$ & $                154.3$ & 26, 42 \\ 
                      FS Lup & SD1 & --- & $ 0.38140$ & $^\textnormal{s} 0.47$ &   ---   & $ 5860$ & $ 5130$ &  M  &  H  & $  58.1$ & $ 5354$ & $   1.8$ & $                 93.7$ & 43 \\ 
                      FU Dra &   C &   W & $ 0.30672$ & $                3.99$ & $ 0.27$ & $ 5823$ & $ 6100$ &  M  &  C  & $   9.9$ & $ 3919$ & $ 281.6$ & $                 85.7$ & 44 \\ 
                      GM Dra &   C &   W & $ 0.33875$ & $^\textnormal{s} 0.18$ & $ 0.23$ & $ 6306$ & $ 6450$ &  M  &  C  & $  18.2$ & $ 5549$ & $ 275.0$ & $                 69.0$ & 38, 45 \\ 
                      GN Boo &   C &   W & $ 0.30160$ & $                3.14$ & $ 0.28$ & $ 5310$ & $ 5068$ &  L  &  H  & $  30.1$ & $ 3717$ & $  92.3$ & $^\textnormal{f}  90.0$ & 46 \\ 
                      GR Vir &   C &   A & $ 0.34697$ & $^\textnormal{s} 0.12$ & $ 0.93$ & $ 6150$ & $ 6554$ &  M  &  C  & $  22.0$ & $ 2460$ & $ 267.0$ & $                  6.0$ & 29, 45 \\ 
                    GSC 1537 &   C &   W & $ 0.31827$ & $                2.65$ & $ 0.08$ & $ 5631$ & $ 5740$ &  M  &  C  & $  15.9$ & $ 4032$ & $ 134.6$ & $                 81.0$ & 47 \\ 
                    GSC 3551 &   C &   A & $ 0.59214$ & $                0.31$ &   ---   & $ 6615$ & $ 6820$ &  L  &  C  & $  22.0$ & $ 4697$ & $ 280.0$ & $^\textnormal{f}  90.0$ & 48 \\ 
                   GSC 03526 &   C &   W & $ 0.29226$ & $                2.85$ & $ 0.18$ & $ 4581$ & $ 4830$ &  M  &  C  & $  22.0$ & $ 3390$ & $ 249.5$ & $                140.3$ & 49 \\ 
                      GU Ori &   C &   A & $ 0.47068$ & $                0.46$ & $ 0.27$ & $ 5940$ & $ 6003$ &  M  &  C  & $  20.9$ & $ 5524$ & $  59.0$ & $^\textnormal{f}  90.0$ & 50 \\ 
                      GW Cep &   C &   W & $ 0.31883$ & $                2.59$ & $ 0.18$ & $ 5800$ & $ 6104$ &  M  &  C  & $  17.8$ & $ 5174$ & $ 282.3$ & $                 69.9$ & 51 \\ 
                      GW Cnc &   C &   W & $ 0.28141$ & $^\textnormal{s} 3.77$ & $ 0.09$ & $ 5649$ & $ 5790$ &  M  &  C  & $  14.6$ & $ 5122$ & $ 113.3$ & $                 85.7$ & 52 \\ 
                      GW Gem & SD2 & --- & $ 0.65944$ & $                0.46$ &   ---   & $ 5004$ & $ 7700$ &  L  &  C  & $  16.2$ & $ 3773$ & $ 112.7$ & $                 80.3$ & 53 \\ 
                      HH Boo &   C &   W & $ 0.31867$ & $^\textnormal{s} 1.70$ & $ 0.10$ & $ 5680$ & $ 5386$ &  L  &  H  & $  16.0$ & $ 4561$ & $  60.0$ & $                 85.1$ & 54 \\ 
                      HS Aqr & SD2 & --- & $ 0.71019$ & $^\textnormal{s} 0.63$ &   ---   & $ 5110$ & $ 6350$ &  L  &  C  & $  43.9$ & $ 3986$ & $ 272.0$ & $                127.0$ & 36, 55 \\ 
                      HV Aqr &   C &   A & $ 0.37445$ & $^\textnormal{s} 0.14$ & $ 0.68$ & $ 6460$ & $ 6599$ &  M  &  C  & $  18.1$ & $ 6169$ & $  37.6$ & $                 66.1$ & 56, 57 \\ 
                      IK Per &   C &   A & $ 0.67603$ & $                0.19$ & $ 0.52$ & $ 9070$ & $ 7470$ &  M  &  H  & $  49.0$ & $ 8163$ & $ 182.7$ & $                143.9$ & 58 \\ 
                      LZ Her &   C &   W & $ 0.33174$ & $                2.59$ & $ 0.18$ & $ 5700$ & $ 6000$ &  M  &  C  & $   9.8$ & $ 5187$ & $ 257.1$ & $                 75.2$ & 59 \\ 
                      NR Peg & SD2 & --- & $ 3.39822$ & $^\textnormal{s} 0.36$ &   ---   & $ 5485$ & $ 4186$ &  M  &  H  & $  23.0$ & $ 4004$ & $ 197.0$ & $                120.0$ & 60 \\ 
                    NSVS 146 &   C &   W & $ 0.39130$ & $                3.51$ & $ 0.30$ & $ 5753$ & $ 5340$ &  L  &  H  & $  19.9$ & $ 3908$ & $  86.4$ & $                 89.1$ & 61 \\ 
                      OO Aql &   C &   A & $ 0.50680$ & $^\textnormal{s} 0.84$ & $ 0.06$ & $ 5700$ & $ 5638$ &  M  &  H  & $  38.9$ & $^\textnormal{f} 3819$ & $  93.2$ & $  16.1$ & 62, 63 \\ 
                      QT UMa &   C &   W & $ 0.47354$ & $                1.71$ &   ---   & $ 6053$ & $ 5497$ &  L  &  H  & $  34.0$ & $ 5750$ & $ 359.0$ & $                112.0$ & 64 \\ 
                      RU UMi & SD2 & --- & $ 0.52493$ & $^\textnormal{s} 0.33$ &   ---   & $ 4630$ & $ 7200$ &  L  &  C  & $  22.6$ & $ 3834$ & $ 329.3$ & $                 50.8$ & 65 \\ 
                      RV CVn &   C &   W & $ 0.26957$ & $                1.74$ & $ 0.10$ & $ 4607$ & $ 4750$ &  M  &  C  & $  12.0$ & $ 3824$ & $  37.0$ & $                132.0$ & 66, 67 \\ 
                      RW CrB & SD2 & --- & $ 0.72641$ & $                0.23$ &   ---   & $ 4448$ & $ 8316$ &  L  &  C  & $  30.3$ & $ 3754$ & $  98.4$ & $^\textnormal{f}  90.0$ & 68 \\ 
                      RZ Tau &   C &   A & $ 0.41567$ & $^\textnormal{s} 0.38$ & $ 0.56$ & $ 7200$ & $ 7300$ &  L  &  C  & $  14.6$ & $ 5832$ & $ 277.2$ & $                 76.5$ & 69 \\ 
                      SS Ari &   C &   W & $ 0.40599$ & $^\textnormal{s} 3.25$ & $ 0.15$ & $ 5860$ & $ 6062$ &  M  &  C  & $  48.2$ & $ 5374$ & $ 283.8$ & $                  3.3$ & 70 \\ 
                      SW Lac &   C &   W & $ 0.32072$ & $^\textnormal{s} 1.25$ & $ 0.31$ & $ 5348$ & $ 5630$ &  M  &  C  & $  44.4$ & $ 4546$ & $ 245.7$ & $                 44.6$ & 71, 72 \\ 
                      TU Boo &   C &   A & $ 0.32428$ & $                0.51$ & $ 0.17$ & $ 5737$ & $ 5800$ &  L  &  C  & $  25.3$ & $ 5008$ & $ 286.2$ & $                 76.7$ & 73 \\ 
                      TY Boo &   C &   W & $ 0.31715$ & $^\textnormal{s} 2.15$ & $ 0.12$ & $ 5469$ & $ 5834$ &  M  &  C  & $  12.9$ & $ 5141$ & $  74.5$ & $^\textnormal{f}  90.0$ & 74 \\ 
                      TY UMa &   C &   W & $ 0.35455$ & $^\textnormal{s} 2.52$ & $ 0.13$ & $ 6250$ & $ 6229$ &  L  &  H  & $  19.7$ & $ 4638$ & $ 135.1$ & $                 28.6$ & 75 \\ 
                      TZ Boo &   C &   A & $ 0.29716$ & $^\textnormal{s} 0.21$ & $ 0.53$ & $ 5890$ & $ 5873$ &  M  &  H  & $   9.2$ & $ 5006$ & $ 134.6$ & $                 90.5$ & 76, 77 \\ 
                       U Peg &   C &   W & $ 0.37478$ & $^\textnormal{s} 3.03$ & $ 0.15$ & $ 5600$ & $ 5800$ &  M  &  C  & $  30.8$ & $ 4088$ & $ 110.4$ & $                137.7$ & 78, 79 \\ 
                   UCAC4 436 &   C &   A & $ 0.36146$ & $                0.40$ & $ 0.08$ & $ 4580$ & $ 4590$ &  L  &  C  & $  42.0$ & $ 4305$ & $   2.0$ & $                 90.0$ & 80 \\ 
                      UW Boo & SD2 & --- & $ 1.00471$ & $                0.41$ &   ---   & $ 4806$ & $ 7880$ &  L  &  C  & $  32.0$ & $ 3845$ & $ 120.0$ & $                 38.8$ & 81 \\ 
                      UZ Leo &   C &   A & $ 0.61806$ & $^\textnormal{s} 0.30$ & $ 0.97$ & $ 6980$ & $ 6830$ &  M  &  H  & $  84.0$ & $ 6819$ & $ 163.9$ & $                111.0$ & 26, 29 \\ 
                   V1073 Cyg &   C &   A & $ 0.78585$ & $^\textnormal{s} 0.30$ & $ 0.17$ & $ 6520$ & $ 6700$ &  L  &  C  & $  50.7$ & $ 5731$ & $ 333.1$ & $                 15.1$ &  7, 82 \\ 
                   V1104 Her &   C &   W & $ 0.22788$ & $                1.60$ & $ 0.15$ & $ 3902$ & $ 4050$ &  M  &  C  & $  30.0$ & $ 3122$ & $  80.0$ & $                166.0$ & 83 \\ 
                   V1123 Tau &   C &   W & $ 0.39994$ & $^\textnormal{s} 0.28$ & $ 0.17$ & $ 5821$ & $ 5920$ &  M  &  C  & $  11.7$ & $ 4564$ & $ 237.6$ & $                 91.6$ & 18, 84 \\ 
                   V1128 Tau &   C &   W & $ 0.30537$ & $^\textnormal{s} 0.53$ & $ 0.13$ & $ 6400$ & $ 6200$ &  L  &  H  & $  24.8$ & $ 5562$ & $  34.0$ & $                 38.0$ &  6, 18 \\ 
                   V1191 Cyg &   C &   W & $ 0.31339$ & $^\textnormal{s} 0.11$ & $ 0.29$ & $ 6215$ & $ 6300$ &  M  &  C  & $   9.9$ & $ 5208$ & $ 166.3$ & $                 90.2$ & 18, 82 \\ 
                   V1695 Aql &   C &   W & $ 0.41278$ & $                0.16$ &   ---   & $ 5500$ & $ 5649$ &  M  &  C  & $  29.5$ & $ 4466$ & $  80.6$ & $                125.0$ & 85 \\ 
                   V1918 Cyg &   C &   A & $ 0.41318$ & $^\textnormal{s} 0.28$ & $ 0.30$ & $ 7300$ & $ 6784$ &  M  &  H  & $  21.1$ & $ 6877$ & $  15.2$ & $                 74.2$ & 86 \\ 
                   V2357 Oph &   C &   W & $ 0.41557$ & $^\textnormal{s} 0.23$ & $ 0.23$ & $ 5640$ & $ 5780$ &  M  &  C  & $  10.0$ & $ 3102$ & $  59.0$ & $                 50.0$ & 28, 87 \\ 
                   V2612 Oph &   C &   A & $ 0.37531$ & $^\textnormal{s} 0.29$ & $ 0.22$ & $ 6280$ & $ 6250$ &  L  &  H  & $  37.9$ & $ 5068$ & $ 210.9$ & $                 57.0$ & 6, 88 \\ 
                    V345 Cas & SD2 & --- & $ 0.68876$ & $                0.50$ &   ---   & $ 4957$ & $ 7400$ &  L  &  C  & $  21.3$ & $ 3866$ & $ 344.3$ & $                 48.6$ & 89 \\ 
                    V376 And &   C &   A & $ 0.79867$ & $^\textnormal{s} 0.30$ & $ 0.55$ & $ 7583$ & $ 8460$ &  L  &  C  & $  48.8$ & $ 6521$ & $ 255.0$ & $                101.2$ & 41, 90 \\ 
                    V388 Cyg & SD1 & --- & $ 0.85904$ & $                0.37$ &   ---   & $ 5543$ & $ 8750$ &  L  &  C  & $  34.1$ & $ 4601$ & $ 223.2$ & $                100.0$ & 91 \\ 
                    V392 Ori & SD2 & --- & $ 0.65928$ & $                0.25$ &   ---   & $ 8300$ & $ 4562$ &  M  &  H  & $  17.5$ & $ 7096$ & $ 177.1$ & $                130.7$ & 92 \\ 
                    V396 Mon &   C &   W & $ 0.39634$ & $                0.39$ & $ 0.05$ & $ 6210$ & $ 5920$ &  L  &  H  & $  11.4$ & $ 5030$ & $ 265.1$ & $                 50.6$ & 93 \\ 
                    V410 Aur &   C &   A & $ 0.36636$ & $^\textnormal{s} 0.14$ & $ 0.72$ & $ 5890$ & $ 5983$ &  M  &  C  & $  12.2$ & $ 4300$ & $ 306.0$ & $                 97.0$ & 28, 87 \\ 
                    V417 Aql &   C &   A & $ 0.37031$ & $^\textnormal{s} 0.36$ & $ 0.31$ & $ 5860$ & $ 6066$ &  M  &  C  & $  10.4$ & $ 1231$ & $ 307.0$ & $^\textnormal{f}  90.0$ & 45, 94 \\ 
                    V502 Oph &   C &   W & $ 0.45339$ & $^\textnormal{s} 0.34$ & $ 0.35$ & $ 5900$ & $ 6140$ &  M  &  C  & $  20.6$ & $ 5133$ & $ 116.4$ & $                102.6$ & 95, 96 \\ 
                    V700 Cyg &   C &   W & $ 0.34005$ & $                1.53$ & $ 0.27$ & $ 5396$ & $ 5770$ &  M  &  C  & $  20.2$ & $ 5099$ & $ 241.8$ & $                 90.0$ & 13 \\ 
                    V737 Per &   C &   A & $ 0.36660$ & $                0.41$ & $ 0.09$ & $ 5624$ & $ 5660$ &  L  &  C  & $  15.6$ & $ 4539$ & $  49.7$ & $                 63.2$ & 97 \\ 
                    V781 Tau &   C &   W & $ 0.34491$ & $^\textnormal{s} 2.47$ & $ 0.21$ & $ 6000$ & $ 5804$ &  L  &  H  & $  35.3$ & $ 3313$ & $  95.9$ & $                  7.7$ & 98 \\ 
                    V789 Her &   C &   W & $ 0.32004$ & $                4.37$ &   ---   & $ 5470$ & $ 4961$ &  L  &  H  & $  22.5$ & $ 3681$ & $ 348.5$ & $                 62.3$ & 99 \\ 
                      VW Boo &   C &   W & $ 0.34232$ & $^\textnormal{s} 0.43$ & $ 0.11$ & $ 5198$ & $ 5560$ &  M  &  C  & $  16.9$ & $ 4002$ & $ 303.7$ & $                 74.5$ & 77, 100 \\ 
                      VW Cep &   C &   W & $ 0.27831$ & $^\textnormal{s} 0.35$ &   ---   & $ 5050$ & $ 5450$ &  M  &  C  & $  32.0$ & $^\textnormal{f} 4000$ & $ 186.0$ & $ 134.0$ & 101 \\ 
                      VW LMi &   C &   A & $ 0.47755$ & $^\textnormal{s} 0.42$ & $ 0.50$ & $ 6180$ & $ 6440$ &  L  &  C  & $  26.0$ & $ 5068$ & $ 181.6$ & $                 27.0$ &  7, 36 \\ 
                      VY Sex &   C &   W & $ 0.44343$ & $^\textnormal{s} 0.31$ & $ 0.22$ & $ 5756$ & $ 5960$ &  M  &  C  & $  19.0$ & $ 4777$ & $ 281.0$ & $                 94.0$ & 28, 87 \\ 
                      WZ Crv & SD2 & --- & $ 1.78878$ & $                0.85$ &   ---   & $ 4220$ & $ 6200$ &  L  &  C  & $  37.2$ & $ 3904$ & $ 178.0$ & $                 90.0$ & 102 \\ 
                      XY Leo &   C &   W & $ 0.28410$ & $^\textnormal{s} 1.65$ & $ 0.02$ & $ 4850$ & $ 4524$ &  L  &  H  & $  40.1$ & $ 3734$ & $ 350.1$ & $                135.3$ & 37, 103 \\ 
                      XY LMi &   C &   A & $ 0.43689$ & $                0.15$ & $ 0.74$ & $ 6093$ & $ 6144$ &  L  &  C  & $  18.6$ & $ 4285$ & $ 135.0$ & $                 82.4$ & 104 \\ 
                      XZ Per & SD2 & --- & $ 1.15163$ & $                0.65$ &   ---   & $ 4628$ & $ 6680$ &  L  &  C  & $  24.0$ & $ 3934$ & $  73.0$ & $                 46.0$ & 105 \\ 
                      YY CrB &   C &   A & $ 0.37656$ & $^\textnormal{s} 0.24$ & $ 0.23$ & $ 6100$ & $ 6499$ &  M  &  C  & $  38.0$ & $ 5673$ & $  26.0$ & $                 35.0$ & 45, 57 \\ 
                      YY Eri &   C &   W & $ 0.32165$ & $                0.47$ &   ---   & $ 5349$ & $ 5600$ &  M  &  C  & $  21.1$ & $ 4119$ & $  64.6$ & $                 83.7$ & 106 \\ 
                      ZZ Aur & SD2 & --- & $ 0.60122$ & $                0.47$ &   ---   & $ 4978$ & $ 7800$ &  L  &  C  & $  19.4$ & $ 3883$ & $  88.8$ & $^\textnormal{f}  90.0$ & 107 \\ 
                     SW J015 &   C &   W & $ 0.21450$ & $                3.13$ & $ 0.15$ & $ 4366$ & $ 4500$ &  M  &  C  & $  16.5$ & $ 3711$ & $  92.2$ & $                 86.5$ & 108 \\ 
                     SW J074 &   C &   A & $ 0.22085$ & $                0.36$ & $ 0.17$ & $ 4400$ & $ 4372$ &  M  &  H  & $  14.6$ & $ 3759$ & $ 107.9$ & $                138.3$ & 109 \\ 
                     SW J075 &   C &   A & $ 0.20917$ & $                0.78$ & $ 0.98$ & $ 3300$ & $ 3224$ &  M  &  H  & $  44.2$ & $ 3049$ & $  13.6$ & $                102.2$ & 110 \\ 
                   2MASS 022 &   C &   W & $ 0.21095$ & $                2.15$ & $ 0.10$ & $ 3759$ & $ 3800$ &  M  &  C  & $  10.2$ & $ 3007$ & $ 269.9$ & $                 86.5$ & 111 \\ 
\hline
\end{longtable}

{
\begin{longtable}[l]{lccrrrrrccrrrrr}
\caption{
Parameters of binary systems with hot spots.\footnotemark[$*$]  \label{HS-binaries}}
  \hline
\multicolumn{1}{c}{Object\footnotemark[$\dag$]} & \multicolumn{1}{c}{Tp} & \multicolumn{1}{c}{STp}	& \multicolumn{1}{c}{$P$}	& \multicolumn{1}{c}{$q$} & \multicolumn{1}{c}{$f$} & \multicolumn{1}{c}{$T_1$} & \multicolumn{1}{c}{$T_2$} & \multicolumn{1}{c}{L/M} & \multicolumn{1}{c}{C/H} & \multicolumn{1}{c}{$\alpha$} 	 & \multicolumn{1}{c}{$T_\textnormal{\scriptsize spot}$} & \multicolumn{1}{c}{lon} & \multicolumn{1}{c}{colat} & \multicolumn{1}{c}{Ref\footnotemark[$\ddag$]} \\
		 &		& 			& \multicolumn{1}{c}{(d)}	& 	  &	   	& \multicolumn{1}{c}{(K)}	& \multicolumn{1}{c}{(K)}	& 	  & 	& \multicolumn{1}{c}{(deg)}	 & \multicolumn{1}{c}{(K)}								 & \multicolumn{1}{c}{(deg)} 	 & \multicolumn{1}{c}{(deg)}	  & 		   \\
\hline
\endfirsthead
\hline
\multicolumn{1}{c}{Object\footnotemark[$\dag$]} & \multicolumn{1}{c}{Tp} & \multicolumn{1}{c}{STp}	& \multicolumn{1}{c}{$P$}	& \multicolumn{1}{c}{$q$} & \multicolumn{1}{c}{$f$} & \multicolumn{1}{c}{$T_1$} & \multicolumn{1}{c}{$T_2$} & \multicolumn{1}{c}{L/M} & \multicolumn{1}{c}{C/H} & \multicolumn{1}{c}{$\alpha$} 	 & \multicolumn{1}{c}{$T_\textnormal{\scriptsize spot}$} & \multicolumn{1}{c}{lon} & \multicolumn{1}{c}{colat} & \multicolumn{1}{c}{Ref\footnotemark[$\ddag$]} \\
		 &		& 			& \multicolumn{1}{c}{(d)}	& 	  &	   	& \multicolumn{1}{c}{(K)}	& \multicolumn{1}{c}{(K)}	& 	  & 	& \multicolumn{1}{c}{(deg)}	 & \multicolumn{1}{c}{(K)}								 & \multicolumn{1}{c}{(deg)} 	 & \multicolumn{1}{c}{(deg)}	  & 		   \\
\hline
\endhead
\hline
\endfoot
\hline
\multicolumn{15}{p{8cm}}{\footnotesize\footnotemark[$*$]The symbols are the same as in Table \ref{CS-binaries}. }\\
\multicolumn{15}{p{8cm}}{\parbox{170mm}{
\footnotesize
\footnotemark[$\dag$]Full names of abbreviation: 
GSC 1537-1557 (GSC 1537), 
1SWASP J075102.16+342405.3 (SW J075), 
1SWASP J155822.10-025604.8 (SW J155). 
}}\\ 
\multicolumn{15}{p{8cm}}{\parbox{170mm}{
\scriptsize
\footnotemark[$\ddag$]References: 
1. \citet{Nelson2010-IBVS5951}; 
2. \citet{Liao2016-ApSS}; 3. \citet{Lazaro2004-MNRAS}; 
4. \citet{Li2001-AJ}; 
5. \citet{Djurasevic2006-PASA}; 6. \citet{Lu1999-AJ}; 
7. \citet{Lee2009-AJ}; 
8. \citet{Gurol2005-NewA}; 9. \citet{Hrivnak1993-ASPC}; 
10. \citet{Yang2012-AJ50}; 
11. \citet{Samec1995-PASP}; 
12. \citet{Zhu2012-AJ}; 
13. \citet{Wang2018-PASJ}; 
14. \citet{Park2013-PASJ}; 15. \citet{Pych2004-AJ}; 
16. \citet{Samec1993-AJ}; 
17. \citet{Lee2010-AJ}; 
18. \citet{Niarchos1992-AA}; 
19. \citet{Yang2013-AJ}; 20. \citet{Rucinski2003-AJ}; 
21. \citet{Djurasevic2013-AJ}; 22. \citet{Rucinski2008-AJ}; 
23. \citet{Hrivnak2006-AJ}; 
24. \citet{Gu2004-AA}; 
25. \citet{Xiang2015-AJ9}; 
26. \citet{Lee2010-AJ898}; 
27. \citet{Gray1997-AJ}; 
28. \citet{Samec2015-AJ}; 
29. \citet{Yang2016-AJ}; 
30. \citet{Zhang2010-PASP}; 
31. \citet{Samec1997-AJ}; 
32. \citet{Djurasevic1998-ApSS}; 33. \citet{Hrivnak1989-ApJ}; 
34. \citet{Djurasevic2011-AA}; 35. \citet{Pribulla2009-AJ137-3655}; 
36. \citet{Williamon2013-PASP}; 
34. \citet{Djurasevic2011-AA}; 37. \citet{Pribulla2009-AJ137-3646}; 
38. \citet{Djurasevic1999-AA}; 39. \citet{Yang2003-AJ}; 
40. \citet{Lu1991-AJ}; 
41. \citet{Zhang2009-AJ}; 
42. \citet{Milone1991-ApJ}; 
43. \citet{Liu2014-RAA}; 
44. \citet{Ekmekci2012-NewA}; 
45. \citet{Lee2014-AJ91}; 
46. \citet{Zhu2009-AJ}; 
47. \citet{Xiang2015-AJ62}; 48. \citet{Lu1986-PASP}; 
49. \citet{Samec2004-AJ}; 
50. \citet{Samec1992-PASP}; 
51. \citet{Ma2018-NewA}; 52. \citet{Hrivnak1995-ApJ}; 
53. \citet{Luo2015-AJ}; 54. \citet{Rucinski1999-AJ}; 
55. \citet{Michaels2017-JAVSO}; 
56. \citet{Jiang2015-PASJ}; 
57. \citet{Djurasevic2016-AJ}; 
}}\\
\endlastfoot
                      AC Boo &   C &   W & $ 0.35243$ & $^\textnormal{s} 3.34$ &   ---   & $ 6241$ & $ 6250$ &  M  &  C  & $  40.0$ & $ 6360$ & $  31.0$ & $                 75.0$ &    1 \\ 
                      AI Dra & SD2 & --- & $ 1.19882$ & $^\textnormal{s} 0.44$ &   ---   & $ 9790$ & $ 6163$ &  M  &  H  & $  13.8$ & $10952$ & $ 166.6$ & $                103.7$ & 2, 3 \\ 
                      AP Aur &   C &   A & $ 0.56937$ & $                0.25$ & $ 0.64$ & $ 9016$ & $ 8703$ &  M  &  H  & $  52.3$ & $ 9737$ & $   8.2$ & $                115.7$ &    4 \\ 
                      AQ Psc &   C &   A & $ 0.47561$ & $^\textnormal{s} 0.23$ & $ 0.23$ & $ 5946$ & $ 6095$ &  L  &  C  & $  36.2$ & $ 6303$ & $ 189.4$ & $                 81.1$ & 5, 6 \\ 
                      AR Boo &   C &   W & $ 0.34487$ & $                2.58$ & $ 0.12$ & $ 5100$ & $ 5382$ &  M  &  C  & $  17.2$ & $ 5268$ & $  39.7$ & $                103.0$ &    7 \\ 
                      AU Ser &   C &   A & $ 0.38650$ & $^\textnormal{s} 0.71$ & $ 0.20$ & $ 5153$ & $ 5495$ &  L  &  C  & $  26.2$ & $ 5369$ & $ 322.3$ & $^\textnormal{f}  90.0$ & 8, 9 \\ 
                      AV Hya & SD2 & --- & $ 0.68340$ & $                0.23$ &   ---   & $ 9400$ & $ 6538$ &  M  &  H  & $  11.5$ & $11327$ & $ 292.2$ & $^\textnormal{f}  90.0$ &   10 \\ 
                      BM UMa &   C &   W & $ 0.27122$ & $                1.85$ & $ 0.17$ & $ 4600$ & $ 4982$ &  L  &  C  & $  25.0$ & $ 5005$ & $ 357.0$ & $^\textnormal{f}  90.0$ &   11 \\ 
                      BS Vul & SD1 & --- & $ 0.47597$ & $                0.34$ &   ---   & $ 4632$ & $ 7000$ &  L  &  C  & $  37.0$ & $ 5466$ & $  96.1$ & $^\textnormal{f}  90.0$ &   12 \\ 
                      BU Vul & SD2 & --- & $ 0.56899$ & $                0.37$ &   ---   & $ 3454$ & $ 5940$ &  L  &  C  & $  31.4$ & $ 5181$ & $ 309.6$ & $                 90.0$ &   13 \\ 
                      BX Dra &   C &   A & $ 0.57902$ & $^\textnormal{s} 0.29$ &   ---   & $ 6980$ & $ 6758$ &  M  &  H  & $  36.5$ & $ 9500$ & $  11.5$ & $                  8.5$ & 14, 15 \\ 
                      CE Leo &   C &   W & $ 0.30343$ & $                0.51$ & $ 0.03$ & $ 4850$ & $ 5111$ &  M  &  C  & $  45.0$ & $ 5020$ & $ 263.5$ & $^\textnormal{f}  90.0$ &   16 \\ 
                      CL Aur & SD2 & --- & $ 1.24438$ & $                0.60$ &   ---   & $ 9420$ & $ 6323$ &  M  &  H  & $  14.1$ & $10598$ & $   2.4$ & $                 71.9$ &   17 \\ 
                      DF Hya &   C &   W & $ 0.33060$ & $                2.36$ & $ 0.12$ & $ 6000$ & $ 5851$ &  L  &  H  & $  25.0$ & $ 7200$ & $ 355.0$ & $^\textnormal{f}  90.0$ &   18 \\ 
                      DZ Psc &   C &   A & $ 0.36613$ & $^\textnormal{s} 0.14$ & $ 0.90$ & $ 6210$ & $ 6124$ &  M  &  H  & $  13.7$ & $ 6583$ & $ 108.2$ & $^\textnormal{f}  90.0$ & 19, 20 \\ 
                      EG Cep & SD2 & --- & $ 0.54462$ & $^\textnormal{s} 0.46$ &   ---   & $ 7850$ & $ 5360$ &  M  &  H  & $  29.8$ & $ 8164$ & $ 351.0$ & $                 90.0$ & 21, 22 \\ 
                      EQ Tau &   C &   A & $ 0.34135$ & $^\textnormal{s} 0.45$ & $ 0.16$ & $ 5800$ & $ 5721$ &  M  &  H  & $  10.8$ & $^\textnormal{f} 6380$ & $ 269.2$ & $^\textnormal{f}  90.0$ &   23 \\ 
                      GR Tau & SD1 & --- & $ 0.42985$ & $                0.22$ &   ---   & $ 3434$ & $ 7500$ &  L  &  C  & $  30.0$ & $ 5151$ & $  44.0$ & $^\textnormal{f}  90.0$ &   24 \\ 
                    GSC 1537 &   C &   W & $ 0.31827$ & $                2.65$ & $ 0.08$ & $ 5740$ & $ 5631$ &  L  &  H  & $  12.5$ & $ 8444$ & $ 300.2$ & $                 84.6$ &   25 \\ 
                      GW Cep &   C &   W & $ 0.31883$ & $                2.59$ & $ 0.18$ & $ 5800$ & $ 6104$ &  M  &  C  & $  10.6$ & $ 6786$ & $   7.3$ & $                 76.2$ &   26 \\ 
                      HL Aur & SD2 & --- & $ 0.62251$ & $                0.84$ &   ---   & $ 6562$ & $ 5351$ &  M  &  H  & $  28.0$ & $ 6667$ & $ 268.0$ & $                 90.0$ &   27 \\ 
                      HR Boo &   C &   W & $ 0.31597$ & $                4.09$ & $ 0.21$ & $ 5743$ & $ 5750$ &  M  &  C  & $  31.5$ & $ 6559$ & $ 181.0$ & $                 53.0$ &   28 \\ 
                      IZ Mon & SD1 & --- & $ 0.77981$ & $                0.39$ &   ---   & $ 4971$ & $ 8500$ &  L  &  C  & $  20.1$ & $ 7109$ & $  88.8$ & $^\textnormal{f}  90.0$ &   29 \\ 
                      KQ Gem & SD1 & --- & $ 0.40799$ & $                0.25$ &   ---   & $ 4641$ & $ 6500$ &  L  &  C  & $  39.3$ & $^\textnormal{f} 6500$ & $ 354.8$ & $^\textnormal{f}  90.0$ &   30 \\ 
                      LP Cep & SD2 & --- & $ 0.69306$ & $                0.79$ &   ---   & $ 6720$ & $ 5043$ &  M  &  H  & $  16.0$ & $ 7594$ & $ 191.0$ & $                106.0$ &   31 \\ 
                      OO Aql &   C &   A & $ 0.50680$ & $^\textnormal{s} 0.84$ & $ 0.07$ & $ 5593$ & $ 5700$ &  L  &  C  & $  15.2$ & $ 7551$ & $   2.6$ & $                 81.4$ & 32, 33 \\ 
                      QX And &   C &   A & $ 0.41217$ & $^\textnormal{s} 0.31$ & $ 0.35$ & $ 6440$ & $ 6420$ &  M  &  H  & $  19.6$ & $ 6569$ & $   2.7$ & $^\textnormal{f}  90.0$ & 34, 35 \\ 
                      RU Eri & SD1 & --- & $ 0.63220$ & $                0.54$ &   ---   & $ 5106$ & $ 6900$ &  L  &  C  & $  14.3$ & $ 6229$ & $  59.0$ & $                 86.0$ &   36 \\ 
                      RW Com &   C &   W & $ 0.23735$ & $^\textnormal{s} 0.47$ & $ 0.06$ & $ 4900$ & $ 4720$ &  L  &  H  & $  14.2$ & $ 5341$ & $  10.1$ & $^\textnormal{f}  90.0$ & 34, 37 \\ 
                      RZ Tau &   C &   A & $ 0.41568$ & $^\textnormal{s} 0.38$ & $ 0.34$ & $ 7085$ & $ 7300$ &  L  &  C  & $  21.8$ & $ 7581$ & $ 354.7$ & $^\textnormal{f}  90.0$ & 38, 39 \\ 
                      SS Ari &   C &   W & $ 0.40599$ & $^\textnormal{s} 0.29$ & $ 0.87$ & $ 5745$ & $ 5950$ &  M  &  C  & $  15.0$ & $ 6320$ & $ 310.0$ & $                 90.0$ &   40 \\ 
                      TX Cnc &   C &   W & $ 0.38288$ & $^\textnormal{s} 2.22$ & $ 0.21$ & $ 6250$ & $ 6537$ &  M  &  C  & $  48.1$ & $ 6531$ & $  91.4$ & $                112.6$ &   41 \\ 
                      TY Boo &   C &   W & $ 0.31715$ & $^\textnormal{s} 2.15$ & $ 0.12$ & $ 5469$ & $ 5834$ &  M  &  C  & $   9.5$ & $ 5852$ & $ 323.1$ & $^\textnormal{f}  90.0$ &   42 \\ 
                   V1799 Ori &   C &   W & $ 0.29030$ & $                1.33$ & $ 0.04$ & $ 4781$ & $ 5000$ &  M  &  C  & $  10.8$ & $ 5737$ & $ 246.0$ & $                 74.5$ &   43 \\ 
                    V357 Peg &   C &   A & $ 0.57845$ & $^\textnormal{s} 0.40$ & $ 0.31$ & $ 7000$ & $ 6687$ &  M  &  H  & $  28.0$ & $ 7595$ & $ 310.0$ & $                 50.0$ & 22, 44 \\ 
                    V407 Peg &   C &   A & $ 0.63688$ & $^\textnormal{s} 0.25$ & $ 0.61$ & $ 6484$ & $ 6980$ &  L  &  C  & $  47.3$ & $ 7333$ & $ 130.9$ & $                 86.0$ &   45 \\ 
                    V473 Cas & SD1 & --- & $ 0.41546$ & $                0.49$ &   ---   & $ 4373$ & $ 5830$ &  L  &  C  & $  35.5$ & $ 4810$ & $  20.6$ & $^\textnormal{f}  90.0$ &   46 \\ 
                    V508 Oph &   C &   A & $ 0.34479$ & $^\textnormal{s} 0.52$ & $ 0.15$ & $ 5893$ & $ 5980$ &  L  &  C  & $  12.5$ & $ 7136$ & $ 281.4$ & $                 72.2$ & 47, 48 \\ 
                    V523 Cas &   C &   W & $ 0.23369$ & $^\textnormal{s} 0.52$ & $ 0.29$ & $ 5104$ & $ 4762$ &  L  &  H  & $  18.0$ & $ 5946$ & $   7.0$ & $                116.0$ &   49 \\ 
                    V865 Cyg &   C &   A & $ 0.36530$ & $                0.45$ & $ 0.17$ & $ 5537$ & $ 5650$ &  L  &  C  & $  16.8$ & $ 7309$ & $  53.0$ & $                 90.0$ &   50 \\ 
                      VZ Psc &   C &   W & $ 0.26126$ & $^\textnormal{s} 0.80$ & $ 0.94$ & $ 4500$ & $ 3949$ &  M  &  H  & $  35.2$ & $ 4776$ & $ 159.7$ & $                 87.2$ & 51, 52 \\ 
                      XZ Leo &   C &   A & $ 0.48774$ & $^\textnormal{s} 0.35$ & $ 0.24$ & $ 7160$ & $ 6981$ &  M  &  H  & $  17.1$ & $ 7726$ & $   5.9$ & $                 89.9$ & 53, 54 \\ 
                      XZ Per & SD2 & --- & $ 1.15163$ & $                0.65$ &   ---   & $ 6680$ & $ 4628$ &  M  &  H  & $  10.0$ & $ 7548$ & $  21.0$ & $                 90.0$ &   55 \\ 
                     SW J075 &   C &   A & $ 0.20917$ & $                0.78$ & $ 0.98$ & $ 3224$ & $ 3300$ &  L  &  C  & $  24.3$ & $ 4198$ & $   8.6$ & $                 74.0$ &   56 \\ 
                     SW J155 &   C &   A & $ 0.26008$ & $                0.65$ & $ 0.07$ & $ 6200$ & $ 5970$ &  M  &  H  & $  18.0$ & $ 6448$ & $   9.0$ & $                 90.0$ &   57 \\ 
\hline
\end{longtable}

\begin{ack}
We thank the anonymous referee for useful comments to improve this paper.
This research was supported by the foreign residency research program of Chukyo University in 2018. 
\end{ack}

\end{document}